\documentclass[a4paper,fleqn,usenatbib]{mnras}

\usepackage{newtxtext,newtxmath}

\usepackage[T1]{fontenc}
\usepackage{ae,aecompl}

\usepackage{graphicx}	
\usepackage{amsmath}	
\usepackage{amssymb}	

\graphicspath{{./Figures/}}

\title[DRAGONS XVI: Thermal Memory of the EoR]{Dark-Ages Reionisation \& Galaxy Formation Simulation XVI: The Thermal Memory of Reionisation}
\author[J. Davies et al.]{
James E. Davies$^{1,2}$\thanks{E-mail: daviesje@student.unimelb.edu.au},
Simon J. Mutch$^{1,2}$, 
Yuxiang Qin$^{3}$, 
Andrei Mesinger$^{3}$,
\newauthor
Gregory B. Poole$^{4}$, 
and J. Stuart B. Wyithe$^{1,2}$\\
$^{1}$School of Physics, The University of Melbourne, Parkville, Victoria 3010, Australia\\
$^{2}$ARC Centre of Excellence for All Sky Astrophysics in 3 Dimensions (ASTRO 3D)\\
$^{3}$Scuola Normale Superiore, Piazza dei Cavalieri 7, I-56126 Pisa, Italy\\
$^{4}$Centre for Astrophysics and Supercomputing, Swinburne University of Technology, PO Box 218, Hawthorn VIC 3122, Australia
}

\date{Accepted XXX. Received YYY; in original form ZZZ}

\pubyear{2019}

\begin{document}
\label{firstpage}
\pagerange{\pageref{firstpage}--\pageref{lastpage}}
\maketitle

\begin{abstract}
Intergalactic medium temperature is a powerful probe of the epoch of reionisation, as information is retained long after reionisation itself. However, mean temperatures are highly degenerate with the timing of reionisation, with the amount heat injected during the epoch, and with the subsequent cooling rates. We post-process a suite of semi-analytic galaxy formation models to characterise how different thermal statistics of the intergalactic medium can be used to constrain reionisation. Temperature is highly correlated with redshift of reionisation for a period of time after the gas is heated. However as the gas cools, thermal memory of reionisation is lost, and a power-law temperature-density relation is formed, $T = T_0(1+\delta)^{1-\gamma}$ with $\gamma \approx 1.5$. Constraining our model against observations of electron optical depth and temperature at mean density, we find that reionisation likely finished at $z_{\rm{reion}} = 6.8 ^{+ 0.5} _{-0.8}$ with a soft spectral slope of $\alpha = 2.8 ^{+ 1.2} _{-1.0}$. By restricting spectral slope to the range $[0.5,2.5]$ motivated by population II synthesis models, reionisation timing is further constrained to $z_{\rm{reion}} = 6.9 ^{+ 0.4} _{-0.5}$. We find that, in the future, the degeneracies between reionisation timing and background spectrum can be broken using the scatter in temperatures and integrated thermal history.
\end{abstract}




\section{Introduction}
Over 13 billion years ago, when the universe was approximately 400,000 years old, it consisted mostly of a neutral atomic gas of hydrogen and helium. Over time the gas began to cool and collapse into the first stars and galaxies. Some of the radiation from these sources was energetic enough to strip electrons from the surrounding atomic gas, ionising it. This period of time, from the birth of the first stars until almost all of the atomic gas in the universe had been reionised, is referred to as the Epoch of Reionisation (EoR). The EoR is the last large-scale cosmic event to be studied in detail, and is of great interest to cosmology as it contains information about the formation processes behind the first galaxies in our universe via their effect on the intergalactic medium (IGM). There are many unanswered questions concerning the EoR, including details of its structure, duration, and effect on subsequent galaxies. 

However, observations of the EoR at optical to near infrared wavelengths are made difficult by the absorption of Lyman alpha photons by neutral hydrogen, which is optically thick to this wavelength even at low concentrations. Since all wavelengths below the Ly$\alpha$ line will eventually redshift to that wavelength, measuring the presence of neutral hydrogen using Ly$\alpha$ optical depth probes the tail end of reionisation \citep{2017MNRAS.465.4838G}, since the absorption saturates around neutral fractions of $x_{HI} \lesssim 10^{-4}$ \citep{2006ARA&A..44..415F}. While saturation varies between sightlines, the strength of this absorption line prohibits direct observations deeper into the EoR. In addition, we can infer the number of free electrons, and hence ionised gas, along a sightline by measuring the Thomson scattering of CMB photons. However this is an integrated measure that cannot distinguish between reionisation histories of different durations.

The reionisation of the IGM is accompanied by a large increase in temperature, to $\sim 2 \times 10^4$ K, followed by cooling on a cosmological timescale \citep{1994MNRAS.266..343M,1997MNRAS.292...27H,2000MNRAS.318..817S,2003ApJ...596....9H,2009ApJ...701...94F,2016MNRAS.460.1885U,2017ApJ...837..106O,2018MNRAS.477.5501K,2018arXiv180104931P,2018arXiv181011683O,2018arXiv181201016G,2019arXiv190704860W}. As a result, the thermal imprint of reionisation will last much longer than reionisation itself. IGM temperature measurements are therefore a potentially powerful way to probe the EoR, as they contain information about the reionisation history of a region that lasts long after it reionises. Modeling the temperature evolution during the EoR allows us to relate various parameters of ionisation history to IGM temperature. Comparing observations of temperature at various redshifts to the model will place constraints on the nature of the EoR and the sources driving it \citep{2002ApJ...567L.103T,2012MNRAS.421.1969R,2014ApJ...788..175L,2017ApJ...847...63O,2019ApJ...872..101B}.

Simulating the thermal history of the IGM is not a new idea, with many authors having studied IGM temperature under various assumptions of the background density and ionisation history. Instantaneous reionisation models \citep{2002ApJ...567L.103T,2003ApJ...596....9H,2016MNRAS.460.1885U}, radiative transfer \citep{2018arXiv180104931P,2018MNRAS.477.5501K}, and inhomogeneous reionisation simulations \citep{2009ApJ...701...94F,2019ApJ...874..154D,2012MNRAS.421.1969R} have all been utilised as a basis for IGM temperature models. Once the density field and ionising background are known, most temperature modelling follows a similar process, where photo-heating, adiabatic cooling due to structure growth and the Hubble flow, along with various cooling processes in the IGM are followed over time. The process used in our model is outlined in \ref{sec:model}.

This work uses the DRAGONS simulation suite, using the density grids from the N-body simulation \textsc{Tiamat} \citep{2016MNRAS.459.3025P} and the ionising flux grids from the semi-analytic galaxy formation model \textsc{Meraxes} \citep{2016MNRAS.462..250M} to determine temperature evolution over time. The ionising flux model in \textsc{Meraxes} captures the inhomogeneous nature of reionisation, allowing us to study the entire thermal structure in our simulation. Importantly, \textsc{Meraxes} directly couples high-redshift galaxy formation to hydrogen reionisation, allowing us to constrain the properties of ionising sources using their effect on the IGM.
Using the temperature to probe the EoR involves comparing our simulated temperature distribution to observations of thermal history, placing constraints on the timing of reionisation, and the sources driving it.

Previous works studying IGM temperature have shown the relationships between the thermal state of the IGM and reionisation. Many of these works connected IGM temperature to statistics of the Lyman alpha forest and offered some measurements and constraints on the nautre of the EoR \citep{2011MNRAS.410.1096B,2013MNRAS.436.1023B,2018arXiv180804367W,2019ApJ...872..101B,2018arXiv181011683O,2019arXiv190704860W}. The hydrodynamic and radiative transfer simulations used in this manner have made it possible to make measurements of the IGM temperature, and draw connections between thermal variables and the EoR. However these simulations are extremely computationally expensive. In order to compare these variables with a wide range of reionisation histories, a faster model is required.
Using the DRAGONS simulation suite, we compare these observations to a wide range of reionisation scenarios in order to statistically constrain the global nature of the EoR.

The paper is structured as follows. The DRAGONS simulations will be briefly described and the post-processing temperature model will be laid out in section \ref{sec:sims}. Overview of the model outputs is given in section \ref{sec:outputs}. Our results, including an investigation of the constraints the IGM thermal history can offer and constraints from current observations, are in sections \ref{sec:results} and \ref{sec:corner} respectively, before concluding in section \ref{sec:conc}.
The cosmology utilised throughout this paper is the flat standard $\Lambda$CDM from \citet{2016A&A...594A..13P} with $\lbrace \Omega_{\rm m},\Omega_{\rm b},\Omega_\Lambda,h,\sigma_8,n_{\rm s} \rbrace = \lbrace 0.308,0.0484,0.692,0.678,0.815,0.968 \rbrace$.
\section{Methodology}\label{sec:sims}

\subsection{Model Inputs}
\textsc{Meraxes} couples early galaxy formation and reionisation in a spatially and temporally dependent way, tracking gas cooling, star formation, and feedback between galaxies and the IGM amongst other processes (see \citet{2016MNRAS.462..250M} for more details). This paper utilises the 100 cMpc \textsc{Tiamat} simulation box \citep{2016MNRAS.459.3025P} containing $2160^3$ particles of mass $2.64 \times 10^6 \rm{M_\odot}$. We use \textsc{GBPTREES} merger trees \citep{2017MNRAS.472.3659P} from redshifts $z=35$ to $z=2$. We use the fiducial parameter balues presented in \citet{2017MNRAS.472.2009Q}, apart from the variations listed in section \ref{sec:sweep}. \textsc{\textsc{Meraxes}} includes a modified version of the excursion-set algorithm \textsc{21cmFAST} \citep{2007ApJ...669..663M} to track the progress of inhomogeneous reionisation in the simulated volume. Emissivity within an ionised bubble of radius $r$ in \textsc{\textsc{Meraxes}} is calculated from the star formation rate within the bubble, $\dot{m}_{*}(r)$
\begin{equation}\label{eq:eps}
\bar{\varepsilon} = \frac{f_{esc}N_\gamma}{\frac{4}{3}\pi r^3(1-0.75Y_{\rm{He}})m_{\rm p}}\dot{m}_{*}(r),
\end{equation}
where $f_{esc}$ is the fraction of ionising photons that escape the host galaxy, $N_\gamma$ is the number of ionising photons produced per stellar baryon, and $m_{\rm p}$ is the proton mass. The specific intensity at the hydrogen ionisation threshold, $J_{21}$, within the region is then computed from the emissivity, \begin{equation}\label{eq:Jto}
\bar{J}_{21} = \frac{(1+z)^2}{4\pi}\lambda _{\rm{mfp}}h_{\rm p}\alpha \bar{\varepsilon},
\end{equation}
where $h_{\rm p}$ is the Planck constant. The comoving mean-free path, $\lambda _{\rm{mfp}}$, is equal to the ionised bubble radius during reionisation, and limited to $30 \rm{cMpc}$ throughout the simulation, due to the maximum scale of the excursion-set algorithm\footnote{The value of $30 \rm{cMpc}$ was chosen to roughly correspond to the mean-free path of ionising photons through an ionisied IGM at $z \sim 6$ \citep{2013MNRAS.432.3340S}.}.
$\alpha$ is the assumed spectral power-law slope, a free parameter in our model where a small value corresponds to a harder UV background, such that
\begin{equation}
J(\nu) = \bar{J}_{21}\left(\frac{\nu}{\nu _{\rm{HI}}}\right)^{-\alpha},
\end{equation}
where $\nu _{\rm{HI}}$ is the ionisation threshold of hydrogen.

\textsc{\textsc{Meraxes}} includes the quasar model detailed in \citet{2017MNRAS.472.2009Q}, where radiation from quasars is included when calculating reionisation structure and feedback, using equations analogous to \ref{eq:eps} and \ref{eq:Jto}, with a spectral slope of $\alpha_q = 1.57$. Grids of the specific intensity from both galaxies and quasars, as well as density grids, are output from \textsc{\textsc{Meraxes}}. As stated in \citet{2017MNRAS.472.2009Q} quasars have a sub-dominant effect on hydrogen reionisation in our model, due to the low number density of these luminous objects.

DRAGONS combines a mass resolution small enough to capture low-mass galaxy formation, a volume large enough to study the structure of reionisation, and an inhomogeneous reionisation model based on galaxy physics. \textsc{Meraxes} can simultaneously reproduce the observed stellar mass function, as well as Thomson scattering optical depth and ionising emissivity measurements with certain parameter choices \citep{2016MNRAS.462..250M}.

\subsection{Temperature Model}\label{sec:model}
In this paper we introduce an IGM temperature model in order to better constrain the EoR within DRAGONS. The model is largely based on the semi-numerical approaches of \citet{SRthesis} and \citet{1997MNRAS.292...27H}. Using the ionising background and density grids from the DRAGONS semi-analytic framework, we calculate the temperature and ionisiation of the IGM within the simulated volume. Non-equilibrium photo-ionisation rates, $\Gamma_i$ and photo-heating rates, $g_i$, are calculated post-ionisation from the specific ionising intensity in \textsc{Meraxes}, $J_{21}$, assuming an optically thin IGM:
\begin{equation}\label{eq:ionrate}
\Gamma _i = \int _{\nu _i} ^{\infty} \frac{4\pi J_\nu}{h_{\rm p} \nu} \sigma _i(\nu)d\nu,
\end{equation}
\begin{equation}\label{eq:heatrate}
g_i = \int _{\nu _i} ^{\infty} \frac{4\pi J_\nu}{\nu}(\nu - \nu _i) \sigma _i(\nu)d\nu,
\end{equation}
where $\sigma_i(\nu)$ is the frequency dependent cross-section taken from \citet{1996ApJ...465..487V}\footnote{We integrate over the assumed power-law spectrum with 100,000 frequency bins between 1 and 4 Ryd. The results are not sensitive to the number of frequency bins, nor is it a limiting factor computationally, as the spectral slope is homogeneous, so this integral only needs to be performed once per model}.
The ionisation state of the IGM is then governed by the following differential equation
\begin{equation}\label{eq:dXdt}
\frac{d\tilde{X}_i}{dt} = -\Gamma _i \tilde{X}_i + \sum _{j,k} \tilde{X}_j\tilde{X}_k R_{jk} \frac{\bar{\rho}_b(1+\delta)}{m_{\rm p}}
\end{equation}
for each species $i,j,k$ $\in$ \{HI,HII,HeI,HeII\} where $R_{jk}$ is the recombination rate of species $j$ and $k$ resulting in $i$, including recombination and collisional ionisation. $\tilde{X}_i$ is defined as $\tilde{X}_i = \frac{n_im_{\rm p}}{(1+\delta)\bar{\rho_b}}$, for the local overdensity, $\delta$, and cosmic mean density, $\rho_b$.

The thermal state of the IGM is governed by the balance between photo-heating, adiabatic cooling under the Hubble flow, recombination cooling, and inverse Compton cooling, as well as changes in local overdensity according to \citet{1997MNRAS.292...27H}

\begin{equation}\label{eq:dTdtfull}
\begin{split}
\frac{dT}{dt} = & \frac{2}{3k_B\sum _i \tilde{X}_i}[G(t) - \Lambda(t,\tilde{X}_i)] - 2HT \\  + & \frac{2T}{3(1+\delta)}\frac{d\delta}{dt} - \frac{T}{\sum _i \tilde{X}_i}\frac{d\sum _i \tilde{X}_i}{dt},
\end{split}
\end{equation}
where $G$ is the total photo-heating rate of all species, $G = \sum _i g_i\tilde{X}_i$, $k_b$ is the Boltzmann constant, and $H$ is the Hubble parameter. The cooling rate, $\Lambda$, takes into account recombinations, collisions, bremsstrahlung, and inverse Compton cooling. We take the rates for these processes from \citet{2015MNRAS.446.3697L}. Photo-ionisation and heating rates are calculated separately for stellar and quasar sources using the optically thin equations \ref{eq:ionrate} and \ref{eq:heatrate}, then added together when solving equations \ref{eq:dXdt} and \ref{eq:dTdtfull}.

Following \citet{SRthesis}, the coupled equations \ref{eq:dXdt} and \ref{eq:dTdtfull} are solved recursively, without assuming ionisation equilibrium, using a first order implicit integration scheme \citep{1997NewA....2..209A,2007MNRAS.374..493B} until they converge to a solution with $| \tilde{X}_{e,k} - \tilde{X}_{e,k+1}|<10^{-6}$ at a given attempt $k$, where we use the electron abundance as our convergence statistic. To improve the efficiency of our code, we adopt a variable timestep, where the timestep length is doubled for the next timestep each time a solution is found, or halved if a convergent solution cannot be found within 100 attempts.

In the same manner, we follow the integrated thermal history via $u$, the total energy injected into the IGM per unit mass via photo-heating. This can be observed in Lyman alpha power spectra, distinct from temperature \citep{2016MNRAS.463.2335N,2019ApJ...872..101B} and can be used to simultaneously measure reionisation timing and amount of photo-heating that exists when only considering mean temperature. The injected energy $u$ is followed by simply integrating the photoheating rate over time. The value of $u$ has been related to the small scale Jeans smoothing of the IGM, as it is dependent on the integrated thermal history throughout the EoR. We cannot calculate the Jeans scale directly in post-processing, but $u$ provides a similar probe into the integrated thermal history throughout the EoR.

In order to achieve the computational speeds required to run the model many times, we track the temperature within a $128^3$ grid with cell side length $\approx 800ckpc$ of the \textsc{Tiamat} $100$ cMpc box; from the redshift of reionisation of each voxel, until $z=4$. The results for this paper use the same 10,000 (approximately 0.5\%) randomly selected voxels in each box, unless otherwise stated, as a sample of the entire volume. As this is a post-processing model, the thermal state of each voxel is treated independently, although their ionising flux intensities and densities are already coupled within \textsc{Meraxes} and \textsc{Tiamat}.

We set the specific intensity above the helium ionisation threshold $\nu_{HeII} = 54.4$ eV to zero, so that there is no reionisation of HeII to HeIII. This is because \textsc{Meraxes} only traces the size of HII bubbles, meaning it does not predict the mean-free path of photons above $\nu_{HeII}$. This will restrict our temperature model to times earlier than HeII reionisation, thought to complete around $z \sim 3$ \citep{2008ApJ...682...14F}. We include HeI reionisation, since we expect helium to be singly ionised at the same time as hydrogen \citep{2003ApJ...586..693W}. When constraining the EoR, we also ignore the outputs of our model below $z=4$, to minimise confusion with the effects of HeII reionisation.

\subsection{Parametrising Reionisation}\label{sec:sweep}
In order to produce and test a large number of thermal and ionisation histories, we vary three parameters within \textsc{Meraxes} and the post-processing temperature model. Escape fraction normalisation and redshift-scaling, as well as background ionising spectral slope.

The temperature of the IGM is sensitive to the timing of reionisation, and the timing of reionisation is heavily dependent on the escape fraction of photons from galaxies. We utilise a redshift-dependent, uniform escape fraction for ionising photons, which was shown by \citet{2016MNRAS.462..250M} to allow the model to match electron optical depth and ionising emissivity observations simultaneously.
\begin{equation}\label{eq:fesc}
f_{esc}=\textrm{min}\left[f_5\left(\frac{1+z}{6}\right)^{\beta},1.0\right]
\end{equation}
We vary $f_{5}$ between 0.03 and 0.12 and $\beta$ between 0 and 2.5 to vary the timing and duration of reionisation in the model. These values were chosen to bracket constraints from electron optical depth measurements, producing reionisation histories that finish between redshifts 5 and 10.

The spectral shape of the ionising background sets the energy injected per photoionisation, which affects the reionisation temperature and the subsequent cooling rate. We model the stellar spectrum between 13.6 and 54.4 eV as a power law (equation \ref{eq:Jto}), with a slope, $\alpha$, between 0.2 and 5. The quasar spectral slope in the same frequency range is fixed at 1.57 and the quasar escape fraction is fixed at 1, as in \citet{2017MNRAS.472.2009Q}.

The range of spectral slopes considered is both broader and softer than those often used in temperature modelling \citep{2016MNRAS.460.1885U,2019ApJ...874..154D}, which are based on Population II stellar synthesis models. This range was chosen to produce at least one thermal history that is consistent with observations for each reionisation history. We investigate these scenarios when studying the correlations between heat injection, reionisation timing and temperature. When placing constraints on the EoR, however, we restrict $\alpha$ to be more consistent with these population synthesis models.
\subsection{Initial Conditions}
For our fiducial model, we start with a 99 per cent ionised (HII and HeII) IGM, with the initial temperature calculated from the UV spectral slope at ionisation and the speed of the ionisation front in \textsc{Meraxes} using fits to radiative transfer simulations, performed by \citet{2019ApJ...874..154D}. If the ionisation front passes through the gas very quickly, the reionisation temperature is decided entirely by the average energy of the ionising photons $\langle E_\gamma \rangle$. \citep{2018MNRAS.477.5501K,1997MNRAS.292...27H}, yielding
\begin{equation}\label{eq:treion}
T_{\textrm{reion}} \approx \frac{1}{3k_b}\langle E_\gamma \rangle.
\end{equation}
However, \citet{2019ApJ...874..154D} found using one dimensional radiative transfer simulations that the ionisation front can pass through slowly enough for collisional cooling within the hot, partially neutral gas to have a large effect on the reionisation temperaure. It was found that the speed of the ionisation front provided the best estimation for reionisation temperature, as the faster the ionisation front passes through the gas, the less time it spends in the hot, semi-neutral state where collisional cooling is efficient.

Ionisation front speeds are calculated in each voxel by finding the distance between ionisation boundaries (where adjacent voxels have different ionisation snapshots) at successive snapshots, and assuming that fronts travel at a constant speed within each 11Myr snapshot. We take the distance between random points in each voxel, representing our uncertainty at the grid resolution. The reionisation temperature is calculated from the front speeds and the spectral slope of the background, using fits provided by \citet{2019ApJ...874..154D}. The distribution of reionisation temperatures in our box is presented in section \ref{sec:treion}, this approach introduces a correlation between ionising flux amplitude, gas density, and reionisation temperature, which further complicates the picture of patchy reionisation.

Since we explore softer spectra than \citet{2019ApJ...874..154D}, initial temperatures for models with $3<\alpha\leq5$ are given by the lowest of our two upper limits; from 1) the temperature at the speed of the ionisation front with $\alpha = 3$ and 2) the maximum temperature given by the spectral slope in equation \ref{eq:treion}. This will slightly overestimate initial temperatures for the slower moving fronts, however softer spectra approach their maximum reionisation temperature at slower speeds, so this effect  will be small.

The total photo-heating energy, $u$, is initialised to the mean excess energy of ionising photons, $E_{\textrm{excess}}$ \citep{2018arXiv180104931P}. Assuming total local absorption of ionising photons,
\begin{equation}\label{eq:excess}
E_{\textrm{excess}} =\frac {\int _{\nu _i} ^{\infty} \frac{4\pi J_\nu}{h_{\rm p} \nu}f_id\nu}{\int _{\nu _i} ^{\infty} \frac{4\pi J_\nu}{h_{\rm p} \nu}h_{\rm p}(\nu - \nu_i)f_id\nu}
\end{equation}
where
\begin{equation}\label{eq:exfactor}
f_i = \frac{n_i\sigma_i}{\sum n_j\sigma_j}
\end{equation}
is a factor that accounts for the preferential absorption of different frequencies by different species, and $j$ sums over HI, HeI and HeII. Regardless of how long it takes the front to move through the gas, the total amount of energy imparted to the gas will depend only on the background spectrum, assuming the number of recombinations and collisional ionisations that occur as the front passes is small, and that all photons with energies between 1 and 4 Ryd are absorbed within the front.

We note that, in contrast to other works, the reionisation temperature is partly decoupled from the average photon energy. This occurs via collisional excitation cooling described above, from the models in \citet{2019ApJ...874..154D}, whereas most previous works assume reionisation occurs very quickly in each region with no cooling in the ionisation front. Using these models lowers Treion without changing the initial $u_0$, as $u_0$ only takes into account energy changes via photo-heating, as defined by \citet{2016MNRAS.463.2335N} and \citet{2019ApJ...872..101B}. As a result, we set the initial $u_0$ to the mean excess energy of the ionising background, described above.
\subsection{Sub-grid Clumping}\label{sec:clump}
Gas on scales below our grid resolution is not homogeneous, and clumping on unresolved scales will increase recombination rates. Increasing the recombination rate will increase temperatures after reionisation by shifting ionisation equilibrium to a more neutral state, allowing more photoionisations to occur, which overcomes the increased cooling rates. Large scale clumping is taken into account via our spatial grid, of voxel length $\sim800$ckpc. We also expect reionisation to erase much of the small scale clumping due to Jeans smoothing. This still leaves some room for subgrid clumping to affect our results, on scales between our grid resolution and the Jeans length.

The increased temperatures due to clumping, while reflective of the total heat inside a voxel, will not represent the wide distribution of temperatures that can exist within the voxel, as it will be dominated by the higher density clumps. In order to compare with higher resolution simulations (see appendix \ref{sec:gridres}), and produce a temperature independent of grid resolution, we set the clumping factor to 1. This implies that we are following the gas at the mean density of each voxel throughout the simulation, rather than each voxel as a whole.

We note that there is an effect due to the subgrid structure on the background spectrum due to spectral filtering. This is where intermediary absorbers harden the ionising background by preferentially absorbing lower frequency photons, as the HI ionisation cross-section scales approximately as $\nu^{-3}$ near the ionisation edge. The amount of hardening is dependent on the structure and clumping of the IGM, as well as the structure or the ionising background. Theoretically this could harden the spectral slope by 3 (see \citet{2009ApJ...703.1416F} appendix D) but measurements of the column density distribution of absorbers suggest a hardening of the spectral slope by approximately 1 \citep{2010ApJ...721.1448S}. However modeling this filtering, as well as the spatially dependent spectrum of the ionising background, is beyond the scope of this work.

\subsection{Model Caveats}\label{sec:assume}
Two approximations are made when evaluating temperature evolution that should be noted. First, equations \ref{eq:dXdt} and \ref{eq:dTdtfull} are applied to the \textsc{Meraxes} grids, rather than individual parcels of gas. Second, we calculate temperatures in post-processing, this assumes independence of each voxel with regards to the temperatures of other voxels, and that any gas influx is at the same temperature as the gas within the voxel. These assumptions can cause some spurious heating or cooling as gas moves through the box, since this will be treated in the same way as structure growth (via the third term of equation \ref{eq:dTdtfull}). Since neighboring voxels tend to have similar ionisation histories and densities, combined with dark matter velocities that are fairly low compared to the voxel sizes, we do not expect this to have a large effect on temperatures.

We tested the effect of gas diffusion on the model by running a model where an extra term was added to equation \ref{eq:dTdtfull} to account for the gas of differing temperature entering the voxel. We keep track of the bulk flow of matter via the Tiamat velocity grids, and conservatively assumed that the temperatures of adjacent regions scaled with the maximum temperature density slope found in our other models, $T \propto (1+\delta)^{0.6}$. Under these assumptions, heating rate differences of order $\sim 10\%$ were observed in high density regions, and changes of order $\sim 1\%$ were observed at mean density. Considering that this model overestimates the heating changes due to gas diffusion (nearby voxels are likely to be at similar temperatures shortly after reionisation), we believe that the independence of voxels is a safe approximation for the purposes of this work.

The independence of voxels within our model also excludes a treatment of recombination emission, since inhomogeneous recombinations are not yet included in \textsc{Meraxes}. The effects of recombination emission are detailed in \citet{2009ApJ...703.1416F} where they find it can contribute up to $10\%$ of the ionising background for hydrogen. The effect of a $10\%$ increase in the photo-ionisation rate will not greatly affect our results (see Appendix \ref{sec:gammachange}). However, recombination emission could soften the background spectrum by $\approx 1$, since the photons from hydrogen recombination will tend to have lower energies than those in the rest of the ionising background. While this would cause a decrease in temperatures, the effect is degenerate with our free parameter for the background spectral slope, so modelling this effect is outside the scope of this work.

We assume that there is negligible heating by $\rm{HeII}$ reionisation before $z=4$ \citep{2016MNRAS.460.1885U}. If Helium reionisation is a highly extended process, then we would be underestimating the temperature, and overestimating the IGM cooling rate for any given reionisation history. With our parameterisation, this would bias our models to earlier reionisation scenarios (due to the flatter temperature gradient), with harder spectral slopes (due to the higher temperatures).

The reionisation model in \textit{\textsc{Meraxes}} assumes local absorption of ionising photons. Photons that would redshift below the ionisation threshold in a full radiative transfer model can still ionise in \textsc{Meraxes}. Since \textsc{Meraxes} is tuned to match the ionising emissivity measurements of \cite{2013MNRAS.436.1023B}, which were calculated based on radiative transfer models, $J_{21}$ could be overestimated by a small amount. This will not have a large effect during Hydrogen reionisation, as the size of HII regions are small enough to make local absorption a good approximation. Furthermore, we find that temperature at all redshifts is insensitive to the value of $J_{21}$ after the region is reionised, as long as it is large enough to maintain a highly ionised IGM (Appendix \ref{sec:gammachange}).

We only vary the escape fraction (equation \ref{eq:fesc}) to produce different reionisation scenarios, while keeping the same source model. This means that we ignore any degeneracy between our escape fraction parameters and source modelling with regards to the IGM thermal state. In particular the constraints we find due to the scatter in temperatures are likely optimistic, as the patchiness of reionisation will have a significant effect on the range of temperatures observed afterwards.

Shock heating of high density regions as they collapse is also not modeled in this work, as we are primarily interested in the diffuse IGM. We only track the increase in temperatures from structure growth on the scale of our grid. This will exclude the $\sim 10^5 K$ gas that exists in hydrodynamic simulations. However, studies of IGM temperature have so far focused primarily on gas at or below the critical density, excluding shock heated gas in their analyses \citep{2018arXiv181011683O}.

\section{IGM Thermal History}\label{sec:outputs}
Using the temperature evolution model described in section 2, we calculated the thermal and ionisation states of the same 10,000 randomly chosen voxels for 2750 \textsc{Meraxes} realisations, in order to obtain a wide range of density and ionising flux histories per model, as well as a wide range of global reionisation and thermal histories. We have also computed the evolution of one full box, to examine the topology of the temperature field. The model parameters used in the illustrative examples within this section are $\lbrace f_{5},\beta,\alpha \rbrace = \lbrace 0.08,1,2 \rbrace$ on the full $128^3$ \textsc{Meraxes} box. This run was chosen as all the correlations between reionisation timing, density and temperature of reionisation are clearly shown in its results. However, this is not our highest likelihood model, based on observational data.
\begin{figure}
  \centering
  \includegraphics[width=\linewidth]{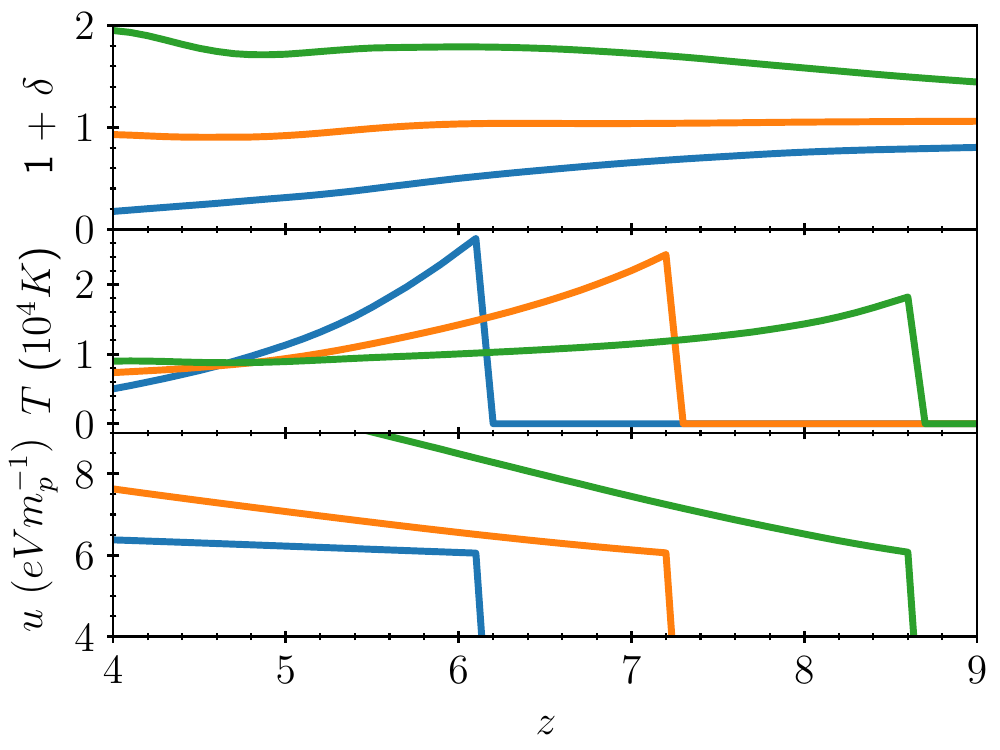}
  \caption{Thermal history of 3 $\sim 800kpc^3$ voxels in \textit{\textsc{Meraxes}}, showing the effect of density history and ionisation timing on the thermal state. Top panel: Density contrast vs redshift. Middle panel: Temperature vs redshift. Bottom panel: Integrated thermal history, showing the total amount of energy injected by photo-ionisations per unit mass.}
\label{fig:single}
\end{figure}
\begin{figure*}
  \centering
  \includegraphics[width=0.92\linewidth]{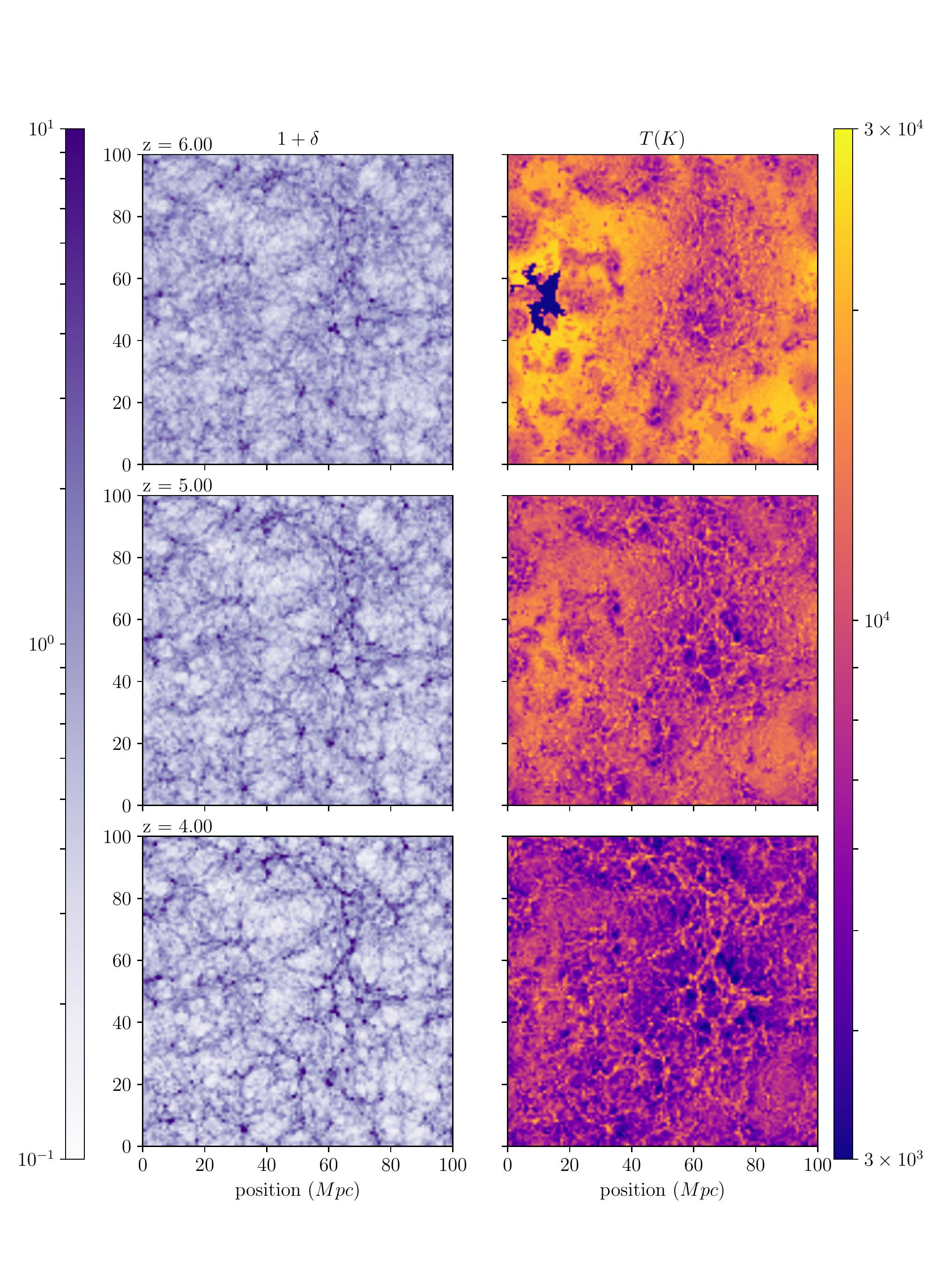}
  \caption{Temperature and overdensity of a full $100 \times 100 \times 0.8 Mpc$ slice of the example model run with parameters $\lbrace f_{5},\beta,\alpha \rbrace = \lbrace 0.08,1,2 \rbrace$ at redshifts 4(bottom), 5(middle) and 6(top). An anti-correlation between temperature and density can be observed towards the end of reionisation $(z=6)$, with very hot voids. This state then decays, forming a tighter positive correlation between T and $\delta$.}
\label{fig:Tslice}
\end{figure*}
\begin{figure*}
  \centering
  \includegraphics[width=\linewidth]{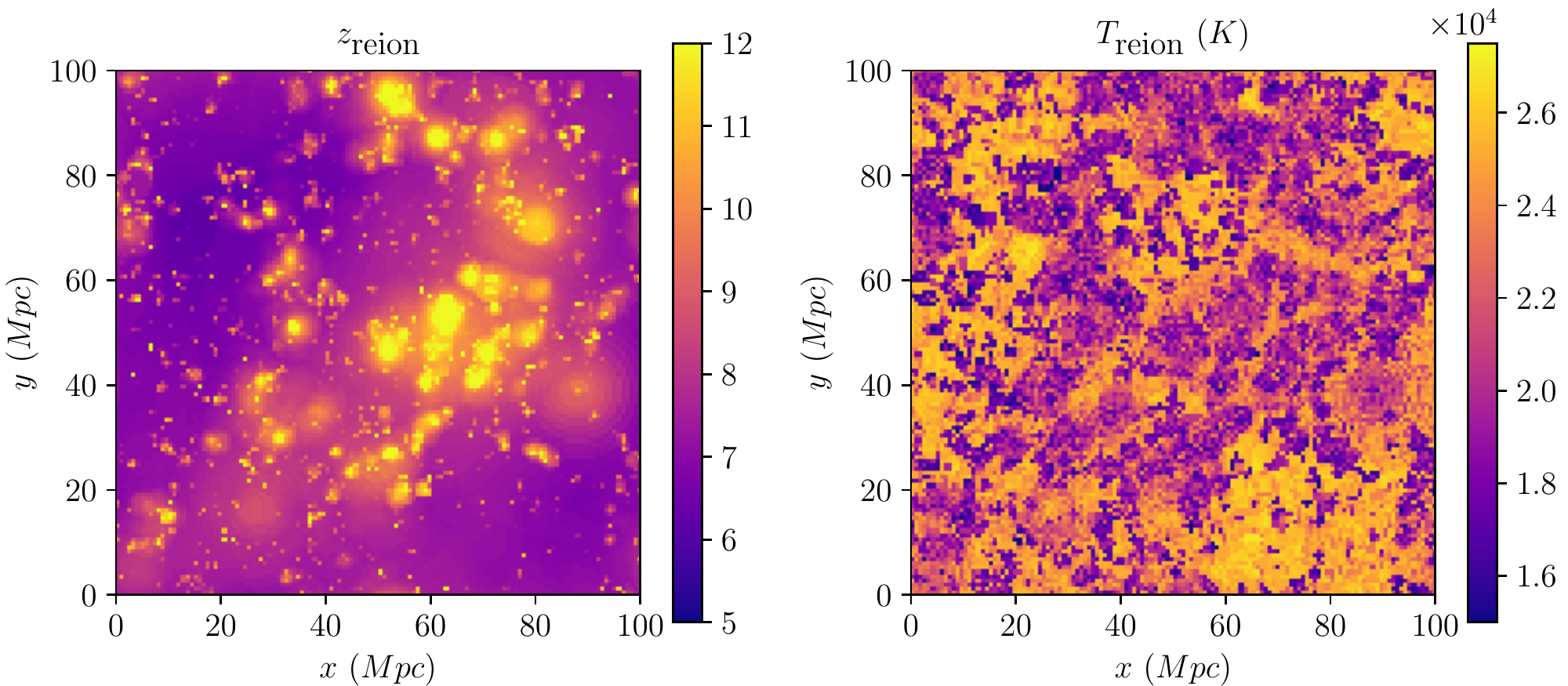}
  \caption{$100 \times 100 \times 0.8 Mpc$ redshift of reionisation slice (left) compared to the temperature just after the region ionises (right), showing an inverse correlation between them, with early reionisation corresponding to low reionsiation temperature. As the ionisation fronts speed up in the cosmic voids, the gas is ionised to a hotter temperature.}
\label{fig:TreS}
\end{figure*}
A few examples of thermal histories output by our model are given in Figure \ref{fig:single}. Each voxel will experience a sudden rise in temperature when it is ionised, dominated by photo-ionisation heating. Afterwards, the voxel cools on a cosmological timescale at a rate determined mainly by the ratio of photo-heating and recombination cooling at ionisation equilibrium, as well as adiabatic cooling and inverse compton cooling off the CMB. Density evolution will modulate the temperature, but the previous density history of a voxel will have little effect on its thermal asymptote \citep{2016MNRAS.456...47M}. However, the previous density history will have a substantial effect on the integrated photo-heating, $u$. Each voxel approches its thermal asymptote within $\Delta z \approx 1-3$ of its reionisation, creating a distribution in temperatures in the whole box dependent on the inhomogeneous reionisation history.
\subsection{Topology}
Figure \ref{fig:Tslice} shows a $128^2$ voxel slice of the temperature model at selected redshifts between 6 and 4. It can be seen that shortly after reionisation, at $z = 6$, the high density regions reach temperatures of $T \approx 2 \times 10^4 K$, as they have recombination rates high enough to allow continuous photo-heating of the gas. The larger voids also reach similar temperatures, as they reionise last due to the ``inside-out" nature of the EoR in our models, and hence have the least time to cool. The coolest regions are the low density regions in close proximity to high density regions, which are reionised early by stars in the nearby dense filament, and cool over time as their recombination rates are not high enough to maintain high levels of photo-heating. Long after reionisation, at $z=4$, the hotter low density regions cool so that temperature is closely correlated with density. The ionisation topology matches that of previous works modeling inhomogeneous reionisation \citep{2008ApJ...689L..81T,2012MNRAS.421.1969R,2018MNRAS.477.5501K,2014ApJ...788..175L,2018arXiv181011683O}. Comparing Figure \ref{fig:Tslice} with the left panel of Figure \ref{fig:TreS}, which show the redshifts of reionisation and temperatures within the same region, we see that shortly after reionisation there is a notable anti-correlation between temperature and redshift of reionisation. Over time this correlation diminishes, and the correlation between temperature and density becomes dominant.
\begin{figure}
  \centering
  \includegraphics[width=\linewidth]{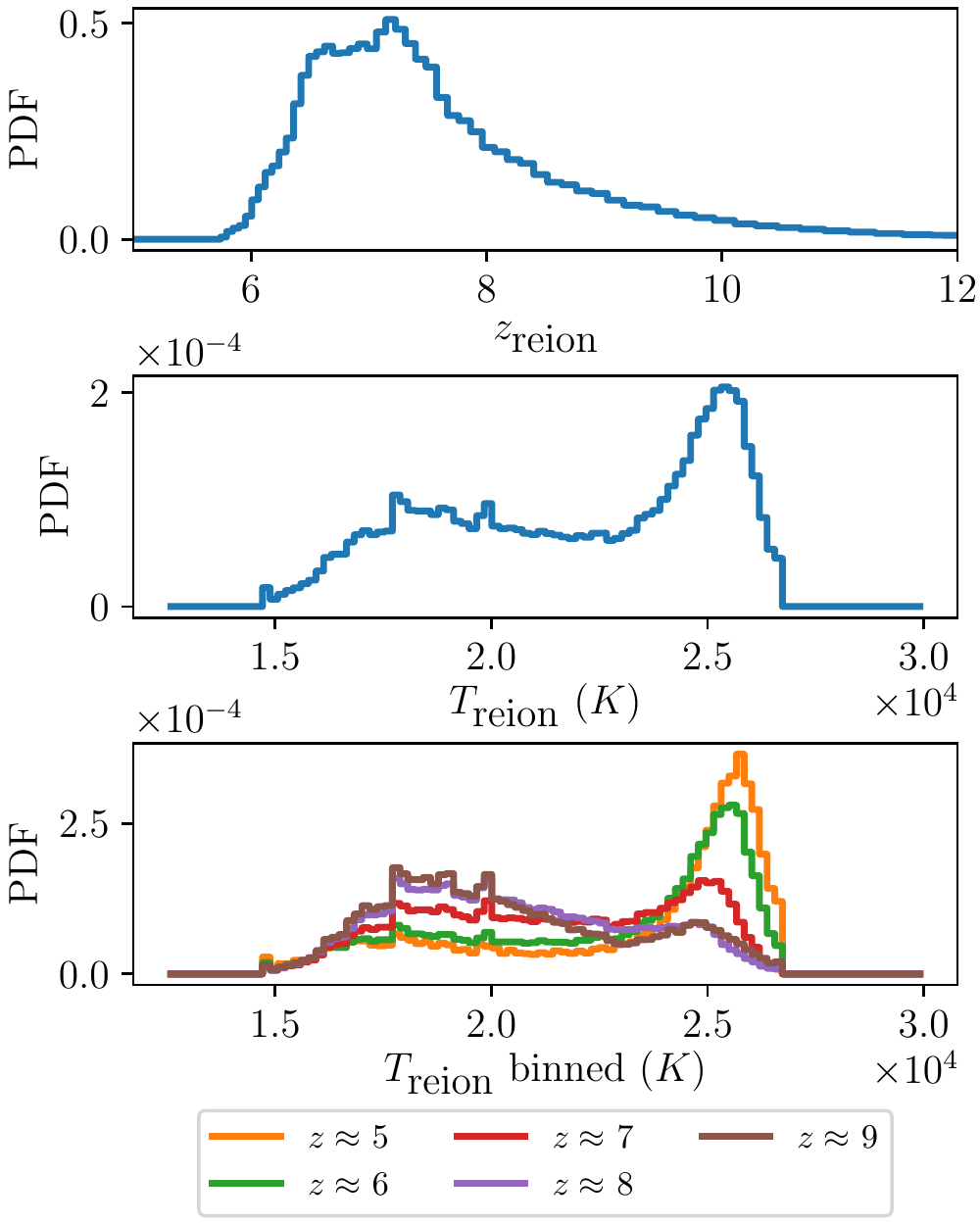}
  \caption{Reionisation temperature distributions in our grid Top panel: redshift of reionisation distribution. Middle: distribution of reionisation temperatures at the time of their reionisation. Bottom: reionisation temperature distributions for voxels reionising at specific redshifts, showing the increasing reionsation temperature as reionisation progesses. Note that these are the temperatures that voxels reionise to for the subset of voxels that reionise near a specific redshift, not the temperatures of all voxels at the same redshift.}
\label{fig:TreH}
\end{figure}

\subsection{Reionisation Temperature}\label{sec:treion}
In this section we present the reionisation temperatures in our fiducial model, using fits for the reionisation temperature provided by \citet{2019ApJ...874..154D}, to convert ionisation front velocities to reionisation temperatures. Using these fits allows us to include the correlations between temperature, ionisation history and the photon background in more detail; this reduces the number of free parameters we need to include.

The temperature to which a region of the IGM is reionised to is inversely correlated with its redshift of reionisation, as shown in Figure \ref{fig:TreS}. This anti-correlation results from the fact that the ionisation fronts speed up as they enter the low-density regions towards the end of the EoR, resulting in higher temperatures because there is less time for collisional cooling to take effect. Differences in simulated reionisation history, and to a lesser extent grid resolution, snapshot cadence, and algorithm to find ionisation front speed result in changes to the distribution. Figure \ref{fig:TreH} shows the distribution of reionisation temperatures, compared to redshift of reionisation. As shown in the top panel, reionisation in this model occurs in the redshift range $6 \lesssim z \lesssim 12$. The middle panel shows the full distribution of temperatures to which regions ionise. We find a similar range of reionisation temperatures as \citet{2019ApJ...874..154D}, between $(15-25) \times 10^4$ K. The bottom panel splits up this distribution into regions that ionised at different times, showing that the temperature of the ionisation fronts increases as reionisation progresses. Lower spatial resolution, as well as our discrete timestep, likely smooths out the extreme ends of the distribution compared to \citet{2019ApJ...874..154D}, and results in a broader distribution of temperatures at any given redshift. However it is difficult to directly compare the distributions due to differences in reionisation history.

The excursion set formalism used to predict reionisation history in \textsc{Meraxes} is likely too crude to predict accurate ionisation front speeds on scales of individual voxels, as the spherical symmetry assumed by \textsc{21cmFAST} effectively averages over many cells during the later stages of reionisation. However, since we recover a similar range of speeds and reionisation temperatures as \citet{2019ApJ...874..154D}, and the reionisation temperature increases as reionisation progresses, we consider this approach to be a valid approximation for testing the statistics of IGM temperature and more accurate than assuming a constant reionisation temperature for all voxels.

\begin{figure*}
  \centering
  \includegraphics[width=\linewidth]{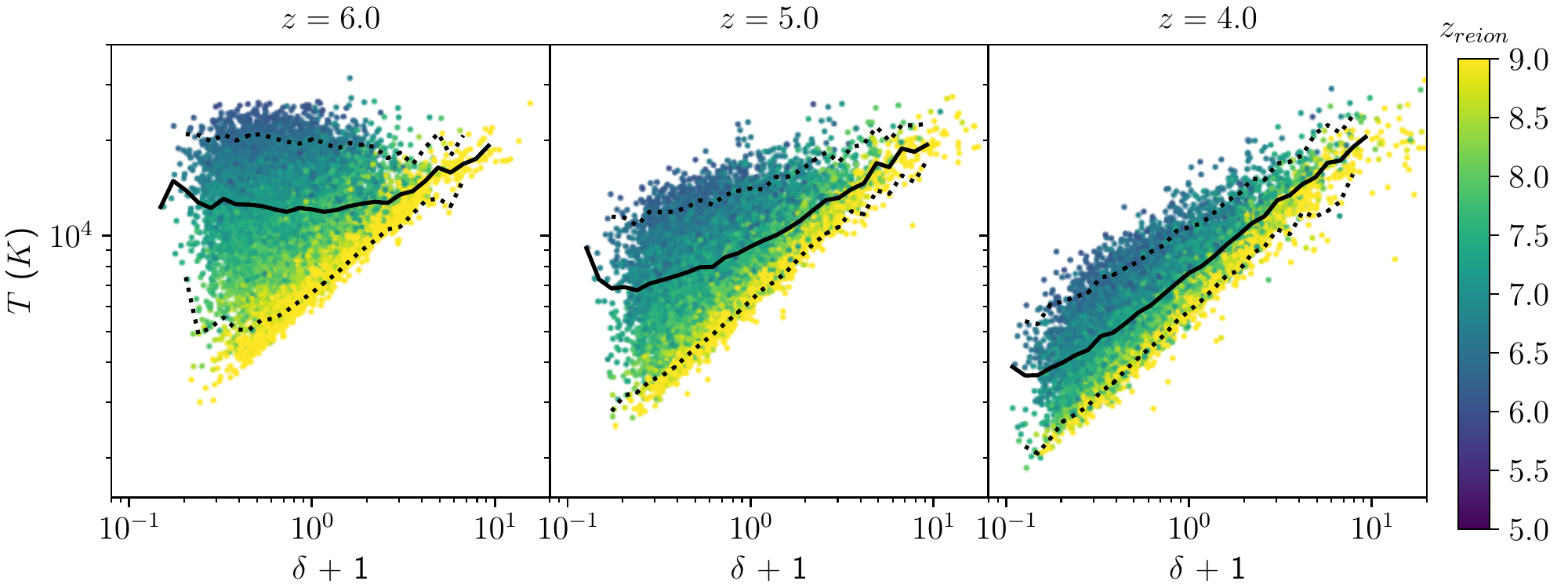}
  \caption{Temperature density relation for 10,000 voxels in our model at redshifts 4 to 6, showing the large scatter in low density regions at early times, followed by a late time power law, with slope approaching $\gamma - 1 \approx 0.5$ by redshift 4 for all densities. The solid black line shows mean temperature at binned denisties, and the dotted black lines show the $5^\textrm{th}$ and $95^\textrm{th}$ quantiles in each bin. The colours of each dot show the redshift at which the voxel ionised.}
\label{fig:tvd}
\end{figure*}

\subsection{Temperature-Density Relation}\label{sec:tdr}
Figure \ref{fig:tvd} shows the temperature-density relation at various redshifts in our fiducial model. The temperature-density relation (TDR) is a powerful probe of the conditions in the IGM. The shape and scatter in the relation at various densities can reveal information about ionisation history and structure of the IGM during the EoR. A power law fit $T=T_0(1+\delta)^{\gamma-1}$ is commonly used when characterising the thermal state of the IGM. However, this is only accurate for regions that ionised homogeneously or long before the measurement is taken. Restricting the study of temperature to a power law misses much of the information in the temperature density relation that can be used to constrain a patchy EoR \citep{2018arXiv181011683O,2019arXiv190704860W}.

There is a wide distribution of temperatures in low density regions shortly after reionisation, as the temperature at low density is highly dependent on the redshift of ionisation at this time. The large scatter lasts long after reionisation itself, halving approximately within $\Delta z \approx 1$ after reionisation in most of our models. Long after reionisation, regions of all densities cool to their asymptotic temperature, and the temperature density relation is well described by a power law $T = T_0(\delta + 1)^{\gamma - 1}$, with a slope approaching $\gamma - 1 \approx 0.6$ by redshift $z=4$. This is consistent with other inhomogeneous reionisation models and analytic calculations of the thermal asymptote \citep{1997MNRAS.292...27H,2016MNRAS.456...47M}. High density regions do not show much variance in their temperatures, as their reionisation temperatures tend to be much closer to their early-reionisation asymptote. As a result high density regions settle much more quickly into the late time power-law. Furthermore, dense regions tend to ionise earlier, so any scatter at the high end of the temperature density relation will likely have disappeared by the time of measurement.

In agreement with other inhomogeneous reionisation models \citep{2008ApJ...689L..81T,2012MNRAS.421.1969R,2018MNRAS.477.5501K}, we find that the large scatter in low density regions shortly after reionisation is highly correlated with the redshift of ionisation of the region (shown in the colours in Figure \ref{fig:tvd}). The size of this scatter, and its correlation with $z_r$, is what causes the changes in the $T-\delta$ slope. The scatter in temperatures can be a powerful probe of the patchiness of reionisation, as it shows differences in reionisation redshift between different regions of the low-density IGM, allowing us to quantify the patchiness of reionisation, as well as estimate how long ago reionisation occurred as the scatter diminishes (see section \ref{sec:mocks}). Once every region has reached its thermal asymptote, and the TDR settles into a power law, the thermal memory of reionisation has been lost, as regions that ionised at different times approach the same temperature, based on their density.

\citet{2008ApJ...689L..81T,2009ApJ...701...94F} and \citet{2012MNRAS.421.1969R} note an inversion of the temperature-density relation, with $\gamma-1\sim-0.2$, at low densities shortly after reionisation, meaning the lowest density regions are hotter on average than mean density regions at the end of the EoR. This inversion is due to the lowest density regions ionising last, hence having less time to cool, as well as ionising to higher temperatures due to faster ionisation front speeds. In the above model, we find this inversion towards the end of reionisation, lasting until $z\approx5$ as there are many hot low-density regions that have recently ionised, and have yet to cool towards the asymptotic power-law relation. The average slope of the TDR is highly dependent on reionisation history and spectral slope, as they alter the strength of the correlations between density, redshift of reionisation, and temperature. As a result, inversion in the TDR is not observed in all of our models, although the slope will always be at a minimum towards the end of the EoR.

\section{Constraining the EoR Using Temperature}\label{sec:results}
\begin{figure}
\includegraphics[width=\linewidth]{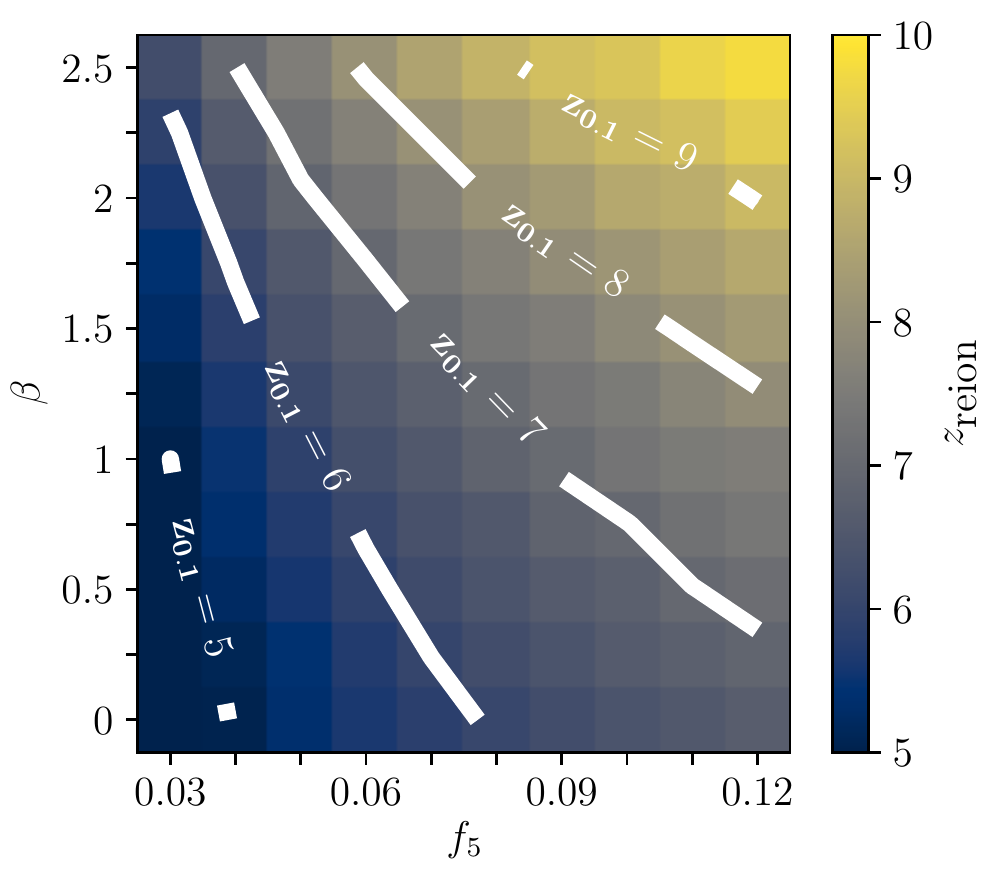}
\caption{Redshift of reionisation of each of our models, showing how $f_5$ and $\beta$ affect reionisation timing. $\alpha$ has a negligible effect on $z_{\rm{reion}}$.}
\label{fig:zre}
\end{figure}
\begin{figure*}
\includegraphics[width=\linewidth]{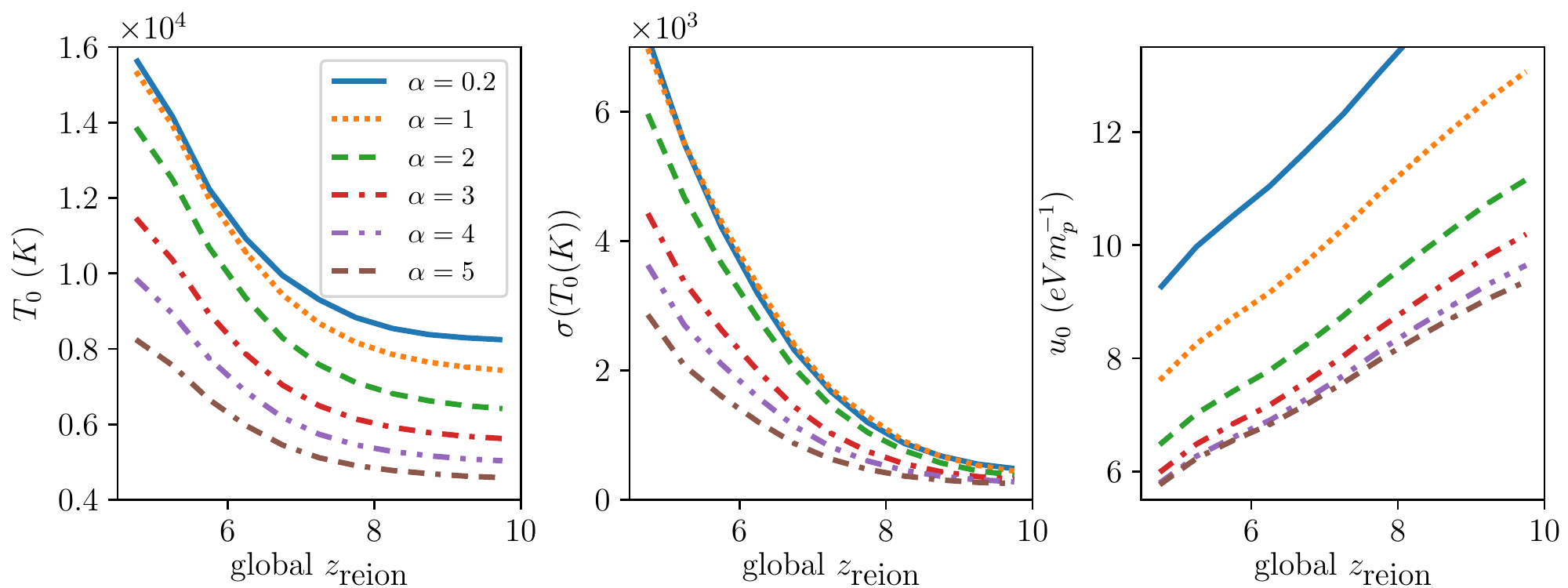}
\caption{IGM thermal state observables at $z=4.5$ versus global redshift of reionisation (when the neutral fraction falls below $0.1$) for different spectral slopes, showing that redshift of reionisation and background spectrum are degenerate in any one observable. Left: Mean temperature at mean density. Middle: Standard deviation in temperature at mean density. Right: Photo-ionisation heat injection per unit mass. The coloured lines in each panel, from top to bottom, represent models with $\alpha = (0.2,1,2,3,4,5)$}
\label{fig:SR1}
\end{figure*}
In this section we investigate how the distribution of temperatures in the IGM can be used to probe reionisation. We have run a suite of realisations of our model, with different escape fractions and background spectral slopes (see section \ref{sec:sweep}), controlling the timing and duration of reionisation, as well as the temperature of reionisation and subsequent cooling rates.

In order to simulate earlier and later reionisation scenarios, we vary the escape fraction normalisation and redshift scaling (equation \ref{eq:fesc}) in \textsc{Meraxes}. A higher (lower) escape fraction results in more (less) ionising photons escaping galaxies, and therefore an earlier (later) reionisation, and a colder (hotter) IGM at a fixed redshift, while thermal memory of reionisation still exists. Models with different escape fractions will tend towards the same temperature at late times, as all regions approach their thermal asymptote. The effects of the escape fraction parameters on global redshift of reionisation (defined as when the global neutral fraction is 10 percent) can be seen in Figure \ref{fig:zre}. Reionisation history also has an effect on reionisation temperatures, by changing the speed of ionisation fronts in the IGM.

Figure \ref{fig:SR1} shows the temperature at mean density, it's standard deviation, and the energy injected from photoionisations at $z=4.5$ versus the model's global redshift of reionisation (where the mass-weighted neutral fraction falls below 0.1) for varying spectral slopes.
As noted in section \ref{sec:tdr}, both the mean and spread of temperature are maximised shortly after the bulk of reionisation occurs. The mean temperature will decrease towards an asymptote dependent on the background spectrum, and the scatter will decrease towards zero. 
A harder spectral slope will impart more energy to the IGM on average per ionising photon, increasing the ratio of the photoheating rate to the ionisation rate. As a result, the thermal asymptote of each voxel becomes hotter. The reionisation temperature also increases, however this can be limited by collisional cooling if the ionisation front passes through the IGM slowly (see section \ref{sec:treion} and \citet{2019ApJ...874..154D}).

Changes in temperature due to reionisation timing and those due to background spectrum are difficult to differentiate using temperature measurements alone. However, mean temperature and scatter in temperature have different correlations with the timing of reionisation and background spectrum, and will evolve at different rates over time. As a result, observations of mean and scatter in IGM temperature at multiple redshifts can be used to break the degeneracy between the timing of reionisation and the background spectrum, offering tighter constraints on the redshift of reionisation.

The integrated photo-ionisation energy at mean density, $u_0$, can be used to further tighten constraints. Unlike temperature, an earlier reionisation increases the photo-ionisation energy at mean density, $u_0$, since residual photo-heating that occurs after reionisation will have been occurring for longer. Because injected energy and temperature have opposite correlations with redshift of reionisation, their observations can be used together to simultaneously constrain the background spectral slope and the timing of reionisation \citep{2016MNRAS.463.2335N,2019ApJ...872..101B}.

\begin{figure}
\includegraphics[width=\linewidth]{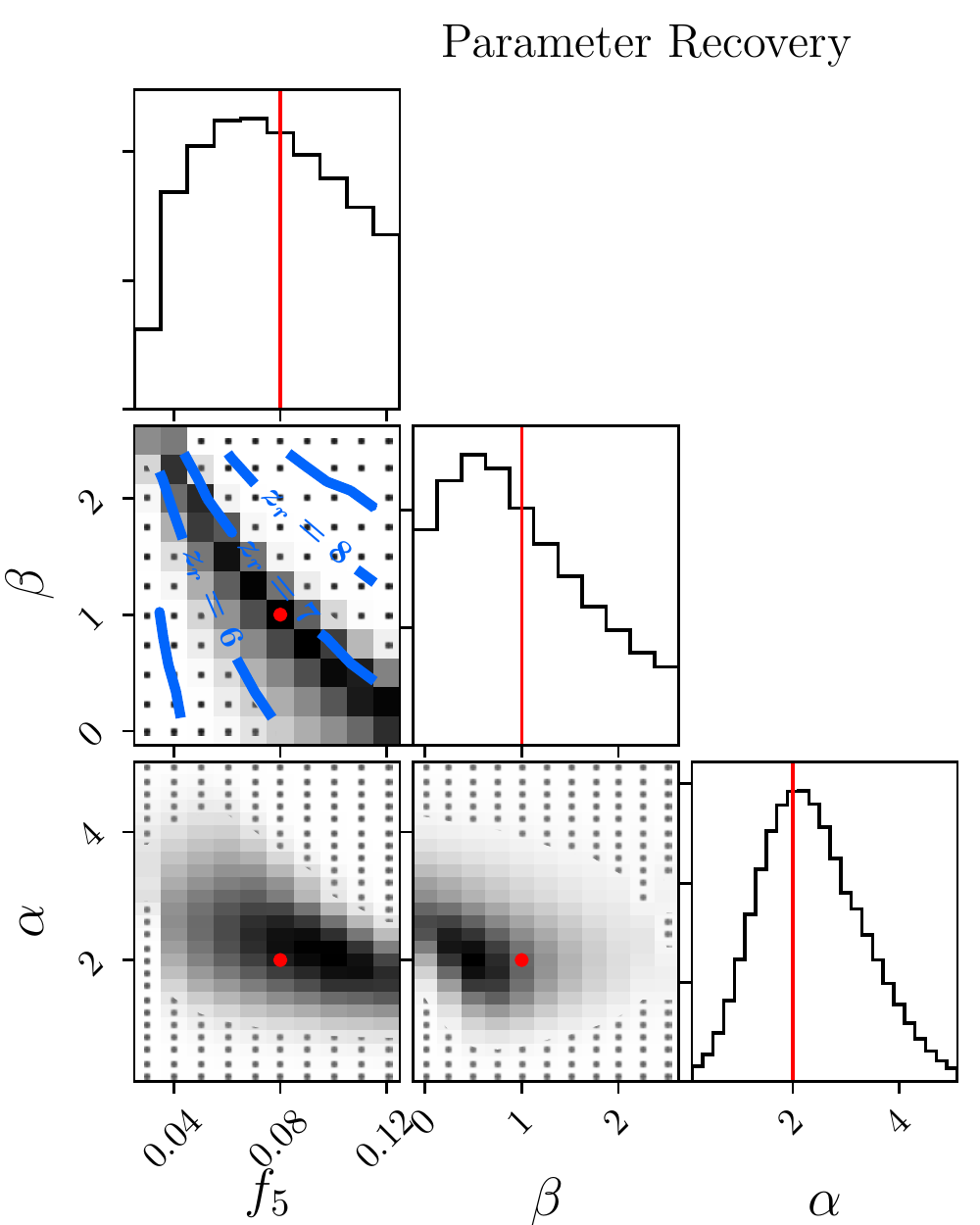}
\caption{Likelihood plots of our suite of models, against mock observations taken from one model, with parameters $\lbrace f_{5},\beta,\alpha \rbrace = \lbrace 0.08,1.00,2.00 \rbrace$, red lines and points show the true model parameters. contours of redshift of reionsiation are plotted in the escape fraction normalisation vs scaling ($f_5$ vs $\beta$) panel, showing our mock constraints on the timing of reionisation.}
\label{fig:zre_cnr}
\end{figure}

\subsection{Mock Observations}\label{sec:mocks}
\begin{figure*}
\includegraphics[width=\linewidth]{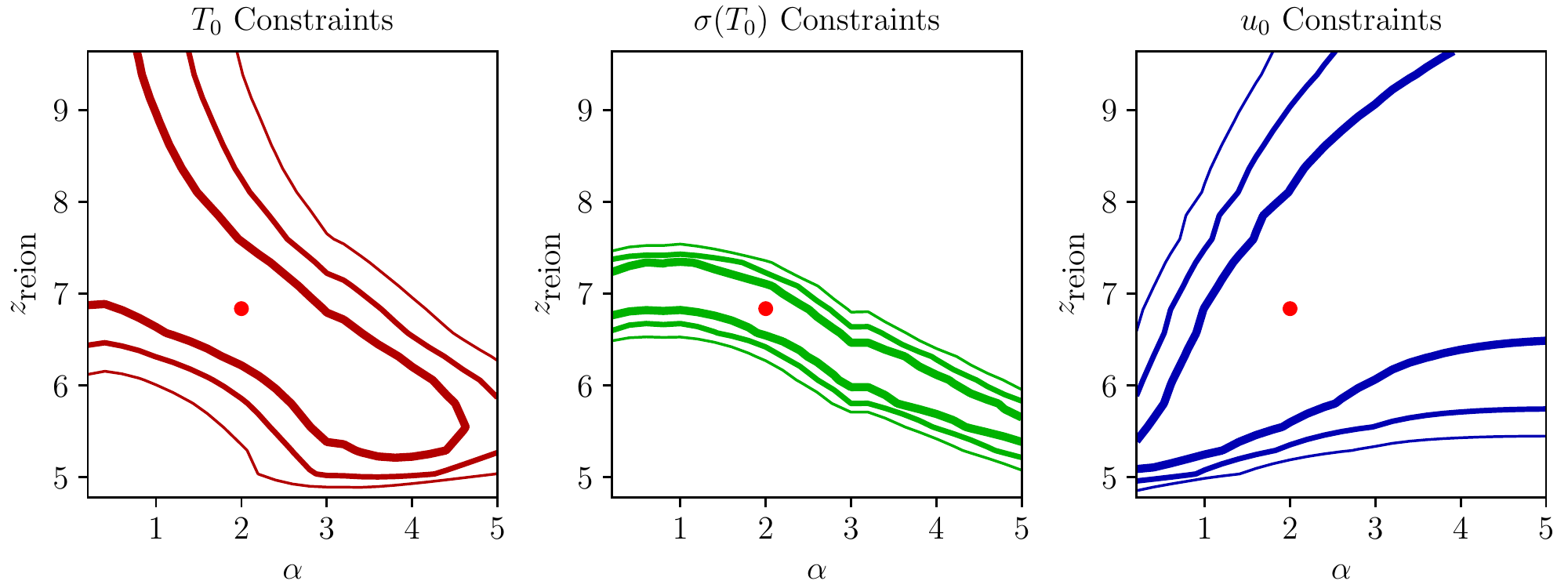}
\caption{Contours of equal likelihood of our models when compared to each observable separately at redshifts 4.5, 5 and 5.5 with 0.125 dex errors, showing the degeneracies of each thermal statistic in $z_{\rm{reion}}$ and $\alpha$. Left panel: mean temperature at mean density. Middle panel: standard deviation of temperature at mean density. Right panel: photo-ionisation energy at mean density. Contour levels, from innermost contour outward, are where the summed square deviation from the mock observation, weighted by variance $\chi^2 = \sum \frac{(\rm{model} - \rm{mock})^2}{\sigma^2}$, is higher than that for the maximum likelihood model by 1, 2.3 and 4. True model values are marked with red dots.}
\label{fig:zhist}
\end{figure*}
In order to explore how temperature observations constrain the EoR, we perform mock observations on our simulation. We create mock observations on the model discussed in section \ref{sec:outputs} of temperature at mean density $T_0$, scatter of temperature at mean density $\sigma(T_0)$, and injected photo-ionisation energy at mean density $u_0$ at redshifts 4.5, 5.0 and 5.5, similar to the redshifts of the temperature measurements from \citet{2019ApJ...872..101B}, which are 4.2,4.6, and 5.0. $T_0$ and $u_0$ are derived from $T$ and $u$ by binning their respective distributions around mean density $\Delta = 0 \pm 0.1$ and averaging. We use 0.125 dex as a $1 \sigma$ measurement error, of similar size to the largest uncertainties given in \citet{2019ApJ...872..101B}. By comparing the mock observations to our series of models, we then estimate the timing of reionisation and background spectral slope, to see how well the input parameters can be recovered. We estimate the likelihood of each model assuming Gaussian errors, based on the true values from one model. The likelihoods of each of our models, compared to these mock observations from our model (with parameters $\lbrace f_{5},\beta,\alpha \rbrace = \lbrace 0.08,1.00,2.00 \rbrace$, and $z_{\rm{reion}} \approx 6.84$), are shown in Figure \ref{fig:zre_cnr}. It is important to note that we do not create mock Lyman alpha spectra, due to the low resolution of our simulations. Here, ``mock observations" refers to the summary statistics of the IGM thermal state, that we generate from one of our models with some assumed error margin. These statistics correspond to the temperature and energy measurements attained from the analysis of Lyman alpha spectra, for example those published in \citet{2011MNRAS.410.1096B,2018arXiv180804367W} and \citet{2019ApJ...872..101B}. 

We recover a strong peak in $\alpha$, and as seen in the contours of the $\beta$ versus $f_5$ panel, the highest likelihood models are those with the same $z_{\rm{reion}}$ as the true model. However, we cannot recover our two escape fraction parameters independently with this sample, as our observables are more sensitive to the timing of reionisation than the duration of it. We will require more precise observations, across a wider range of redshifts to begin to distinguish these scenarios using the scatter in temperature. A more extended reionisation will have a wider distribution of temperatures, from a wider distribution of reionisation times.
\begin{figure*}
\includegraphics[width=\linewidth]{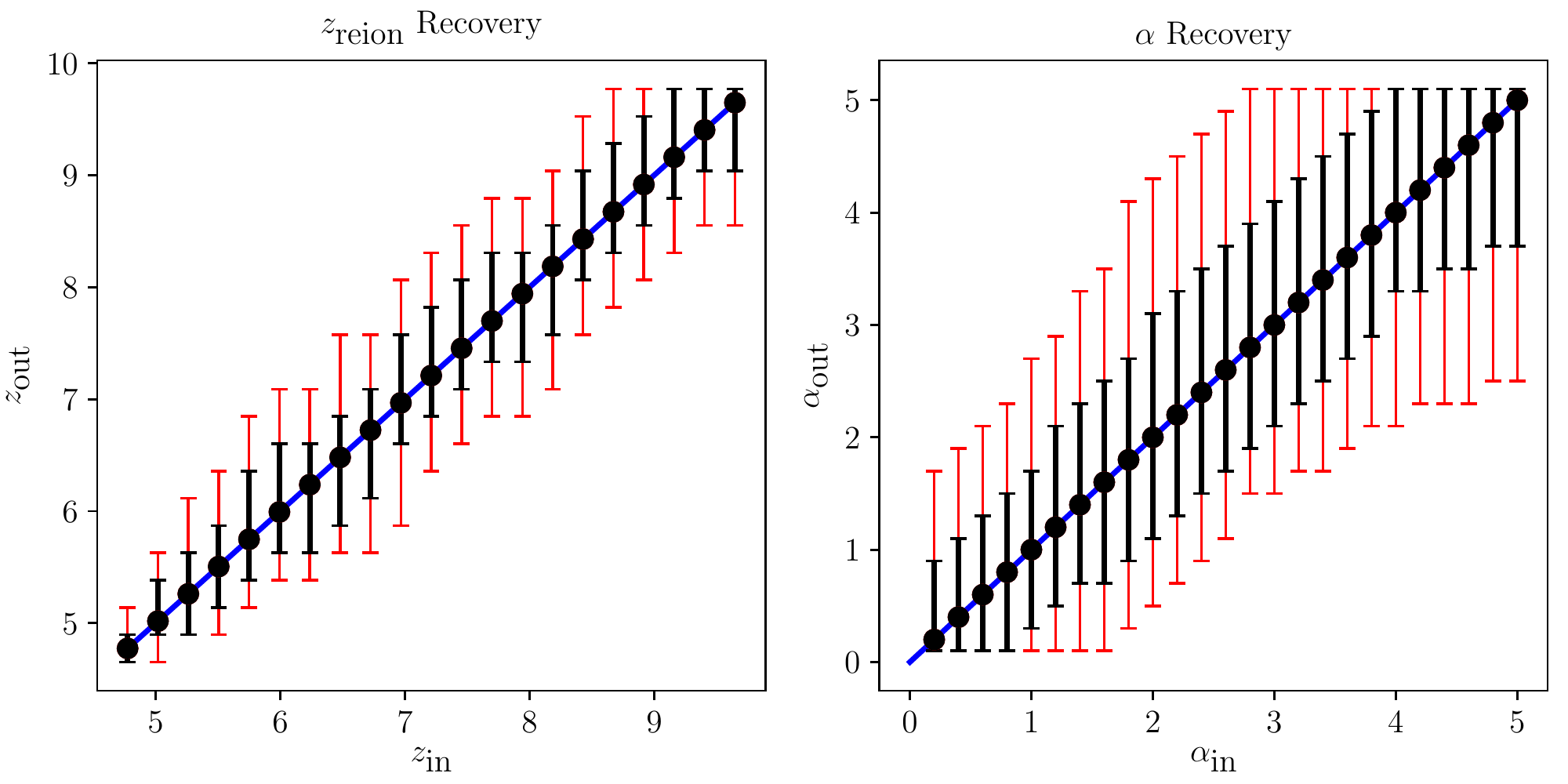}
\caption{Parameter recovery in all of our models. The x axes show the input model parameters and the y axes show the parameter likelihood distribution when the entire set of model runs are compared to the true thermal variables, assuming gaussian errors of 0.125 dex, of models with those parameters. Left: redshift of reionisation, right: background spectral slope. Black errorbars show 68\% probability limits, red errorbars show 95\% probability limits. The blue line $z_{in} = z_{out}$, and the maximum likelihood models (black points), are naturally the same in parameter recovery.}
\label{fig:zre_io}
\end{figure*}

\begin{figure}
\includegraphics[width=\linewidth]{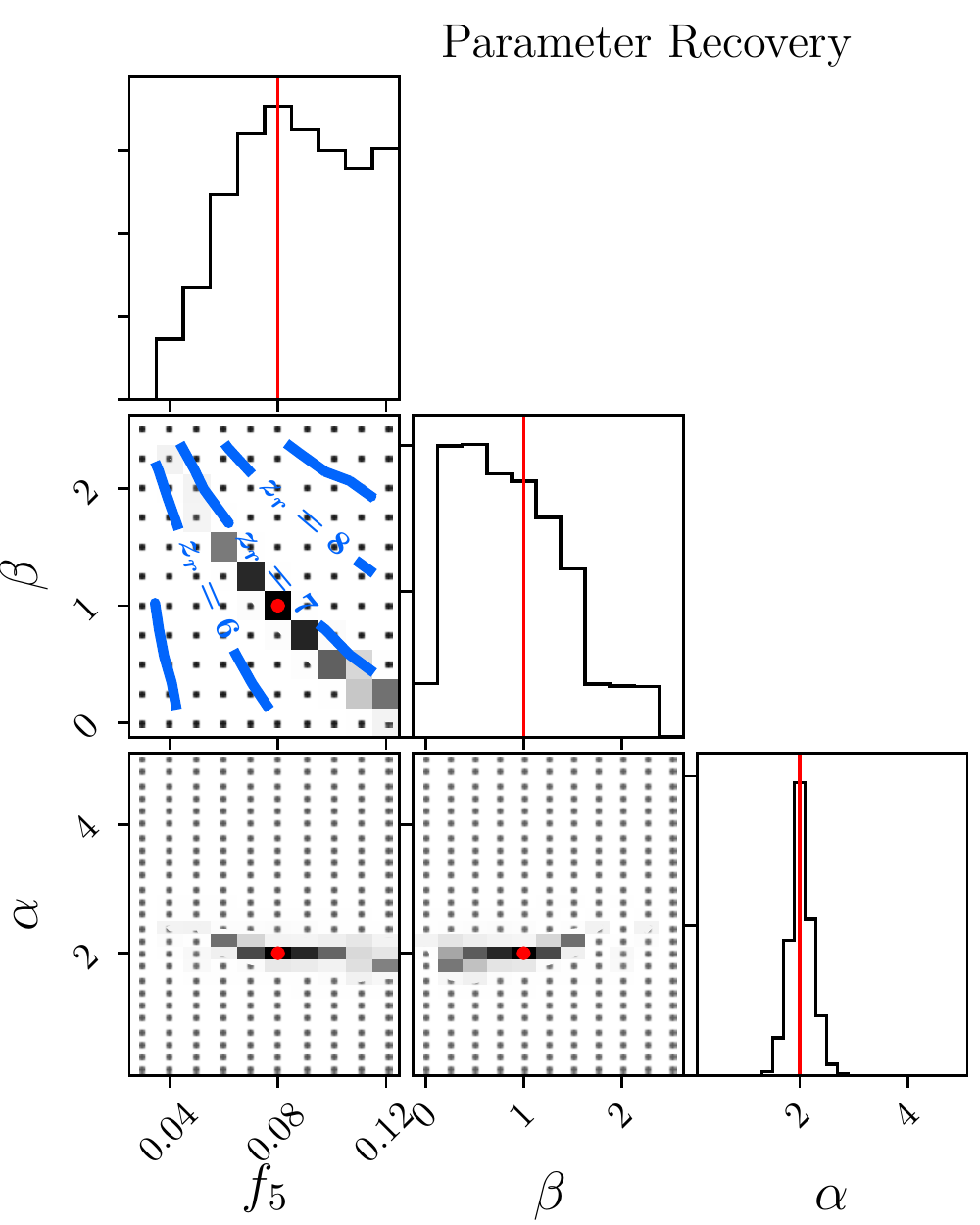}
\caption{Same parameter recovery as Figure \ref{fig:zre_cnr}, based on an optimistic 0.05 dex errors in 9 redshift bins between redshifts 4 and 6. Showing much tighter constraints on model parameters, we are closer to recovering both escape fraction normalisation and redshift scaling.}
\label{fig:zre_low}
\end{figure}
We next present the constraints from each observable separately. The different correlations of temperature at mean density, its scatter, and injected energy with redshift of reionisation and background spectral slope allow us to constrain these reionisation parameters simultaneously. A similar result was demonstrated by \citet{2019ApJ...872..101B}, using the integrated thermal history to break the degeneracy between reionisation timing and initial temperatures. Figure \ref{fig:zhist} illustrates this for the $\lbrace f_{5},\beta,\alpha \rbrace = \lbrace 0.08,1.00,2.00 \rbrace$ model, showing the constraints from mock observations of mean temperature, scatter and photo-heating separately on the redshift of reionisation and background spectrum. While each observable alone is degenerate between the timing of reionisation and the background spectrum, their degeneracies differ in magnitude and direction, allowing us to perform this analysis.

We have applied this analysis to each of our models and present the precision achievable for recovery of reionisation redshift and spectral slope in Figure \ref{fig:zre_io}. We can recover the redshift of reionisation in our model within $\Delta z \approx 1$ and spectral slopes within $\Delta \alpha \approx 2$ to 95\% confidence, using an error margin of 0.125 dex for temperature and injected energy observations.

\subsection{Future Observations}
\begin{figure*}
\includegraphics[width=\linewidth]{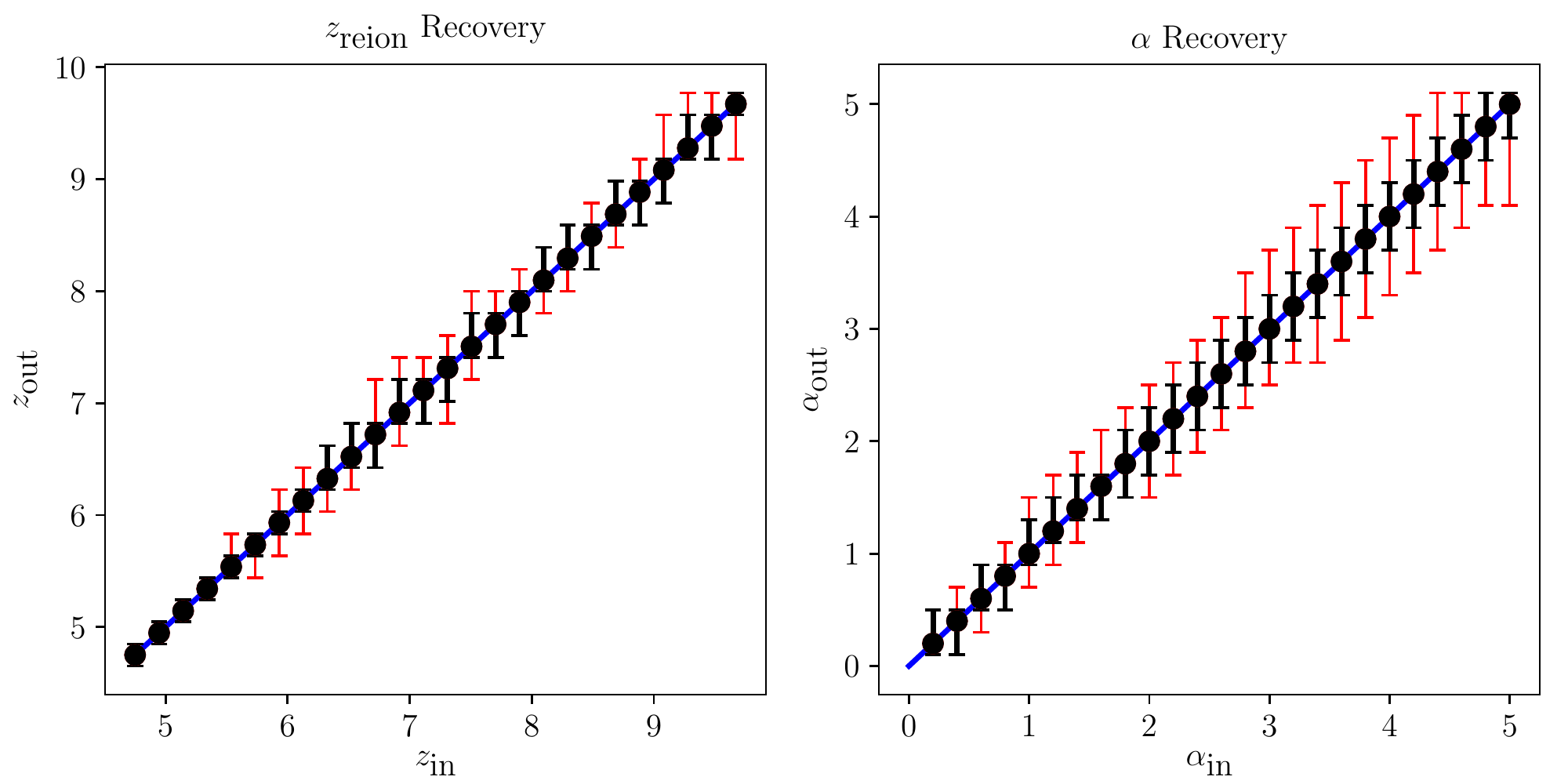}
\caption{Similar to Figure \ref{fig:zre_io}, based on the same optimistic mock dataset as Figure \ref{fig:zre_low}. Parameter recovery with these uncertainties shows a much tighter constraint on redshift of reionisation and background spectra.}
\label{fig:zio_low}
\end{figure*}

As discussed in section \ref{sec:mocks}, our mock observations with errors corresponding to existing datasets were unable to independently recover escape fraction normalisation and scaling. We now examine a much more optimistic dataset from possible future observations, where temperatures at mean density, scatters, and injected energies are known within a $1\sigma$ error of 0.05 dex at 9 evenly spaced redshifts between 4 and 6. We show the results for this case in Figure \ref{fig:zre_low} for the parameter set $\lbrace f_{5},\beta,\alpha \rbrace = \lbrace 0.08,1.00,2.00 \rbrace$ and Figure \ref{fig:zio_low} for all models. The recovery of parameters is much more precise in this scenario. With such an extensive dataset we can recover the redshift of reionisation within $\Delta z \approx 0.5$ and background spectral slope within $\Delta \alpha \approx 0.5 also$ to 95\% confidence\footnote{We note that our model grid is not sufficiently fine to resolve details of the 2d parameter space, however the likelihood values illustrate the improved constraints}. This example is also closer to recovering the escape fraction normalisation and scaling independently. These mock measurements of the scatter in temperatures disfavour reionisation scenarios that are too extended or too sudden, however there is still a significant degeneracy between these two recovered parameters. Considering that this is a very optimistic dataset, recovering the duration of reionisation from its thermal state will likely require a more detailed analysis.

\section{Constraints Using Existing Observations}\label{sec:corner}
\begin{figure}
\includegraphics[width=\linewidth]{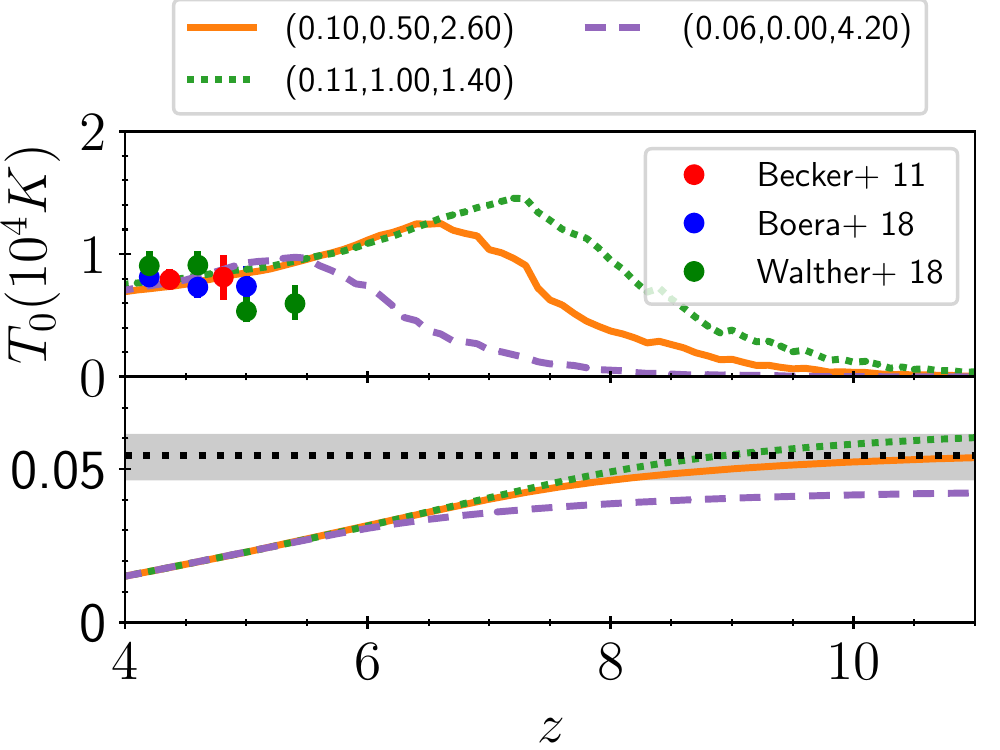}
\caption{Observations used for our parameter constraints, plotted against a selection of models. Model parameters are shown in the legend in the form $\lbrace f_{5}, \beta, \alpha \rbrace$. The degeneracy between reionisation timing and background spectrum is shown here, as models with very different reionisation histories have approximately the same $T_0$ in the interval $4<z<5$.} 
\label{fig:bestrun}
\end{figure}

In the last section, we demonstrated how IGM thermal state observables constrain the nature of the EoR using mock observations of our suite of models.
In this section, we apply this process to recent observations of the thermal and ionisation history of the universe, in order to constrain our model parameters. We constrain against measurements of temperature at mean density from \citet{2019ApJ...872..101B}, and electron optical depth measurements from \citet{2018arXiv180706209P}. Temperature measurements from \citet{2011MNRAS.410.1096B} and \citet{2018arXiv180804367W} were also consdidered, but model likelihoods on these datasets will not be presented in this paper. The former dataset produces similar but looser constraints on reionisation, whereas we are unable to match temperatures from the latter dataset due to the sudden increase in temperatures at $z=5$. Figure \ref{fig:bestrun} shows the observations utilised, and maximum likelihood models based on each temperature dataset. Electron Optical depth measurements place constraints on the redshift of reionisation\footnote{Electron optical depth can also be used to constrain both the timing and duration of reionisation, though not independently \citep{2017MNRAS.465.4838G}. However, we only constrain $z_{\textrm{reion}}$ in this work}, while temperature measurements constrain both the timing of reionisation and the background slope. 

We restrict our attention to temperature at mean density, $T_0$, and electron optical depth, $\tau_e$. No measurements currently exist for scatter in temperatures, so we are unable to use this observable for our EoR constraints, as in our mock examples. Regarding $u_0$, we are unable to reliably relate the integrated thermal history in the optically thin models that produce this measurement in \citet{2019ApJ...872..101B}, with the thermal histories in our patchy reionisation models. A patchy reionisation effectively has a much harder spectrum within ionisation fronts, reducing the effective spectral slope by approximately 3 due to all ionising photons being absorbed within the front\footnote{Assuming the number of recombinations in the ionisation front is small, the photoheating energy at ionisation is set by $E_{\rm{excess}}$ (equation \ref{eq:excess}, see section \ref{sec:treion}) regardless of front speed. In a optically thin reionisation, both integrands would have a factor of the ionisation cross-section $\sigma_{\rm{HI}}$, which scales approximately as $\nu^{-3}$, softening the effective UV spectrum}. This creates a hotter reionisation, but most importantly, permanently offsets the injected energy by a certain amount compared to an optically thin model, because $u_0$ is a time-integrated statistic. Since we cannot model how $u_0$ affects the small scale Lyman alpha power spectrum in our patchy reionisation models, we do not include $u_0$ in our fiducial constraints.

We have ignored temperature measurements at redshifts $z<4$ to minimise confusion resulting from the beginning of HeII reionisation. If a substantial amount of HeII ionisation has occurred at the redshift of measurement, this could be confused with a hotter post HI reionisation IGM with a flatter or positive evolution. This would result in a bias towards models where the overall cooling rate is suppressed, either due to a harder ionising spectrum or earlier reionisation, where the gas is closer to its thermal asymptote and would show a flatter evolution.

\begin{figure*}
\includegraphics[width=\linewidth]{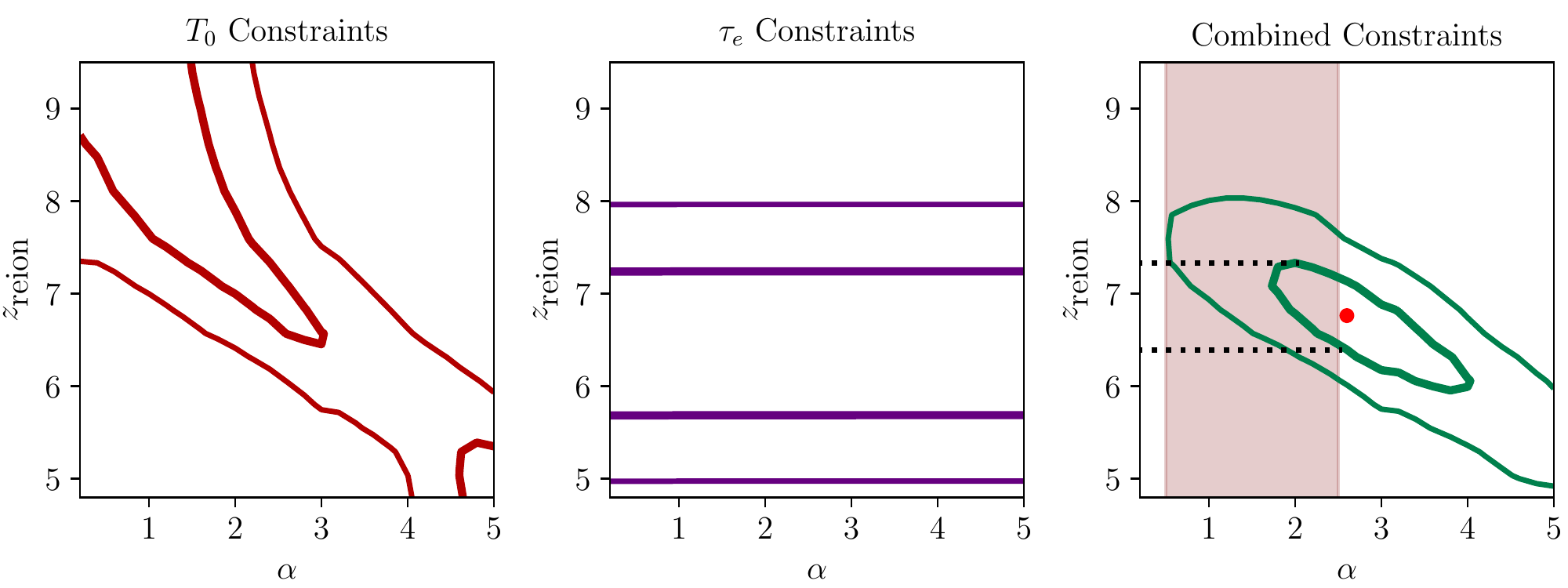}
\caption{Contours of equal likelihood provided by each observable in redshift of reionisation and background spectral slope. Left panel: mean temperature at mean density. Middle panel: Electron optical depth. Right panel: Combined constraints.  Contour levels, from innermost contour outward, are where the summed square deviation from the true value, weighted by variance $\chi^2 = \sum \frac{(\rm{model} - \rm{obs})^2}{\sigma_{obs}^2}$, is higher than that for the maximum likelihood model by 1, 2.3 and 4. For the right panel, these contours, from innermost outward, relate to the projected 1-dimensional 68\% and 95\% confidence intervals for each parameter. In the right panel, the red-shaded region shows the approximate range of background spectral slopes $\alpha$ found by \citet{2019ApJ...874..154D} using various stellar population synthesis models. Horizontal dotted lines show the $1 \sigma$ range of $z_{\rm{reion}}$ within this range of $\alpha$, and the red dot shows the maximum likelihood model.}
\label{fig:rhist}
\end{figure*}

Given the strong degeneracy between escape fraction parameters, we instead show constraints on $\alpha$ and the redshift of reionisation, $z_{\rm{reion}}$. Figure \ref{fig:rhist} shows our constraints on these parameters based on the measurements from \citet{2019ApJ...872..101B} and \citet{2018arXiv180706209P}. We find a strong degeneracy between $z_{\rm{reion}}$ and $\alpha$ where $T_0$ alone is considered. Using $\tau_e$ to break the degeneracy results in a late reionisation $z_{\rm{reion}} = 6.8 ^{+ 0.5} _{-0.8}$ and a soft ionising background $\alpha = 2.8 ^{+1.2} _{-1.0}$.

In order to match temperature observations at $z>4$ as well as observations of electron optical depth, our simulations favour a softer UV spectrum than assumed in other works. This may result from tension between the observations, with temperature observations favouring an earlier reionisation than electron optical depth measurements. This is certainly the case when we restrict the background spectral slope to ranges supported by stellar population synthesis models, with $0.5 < \alpha < 2.5$, given by the red shaded region in figure \ref{fig:rhist} (see \citet{2016MNRAS.460.1885U} and \citet{2019ApJ...874..154D} for discussions of possible background spectral slopes) which tightens our reionisation timing constraints to $z_{\rm{reion}} = 6.9 ^{+ 0.4} _{-0.5}$. However many models without this restriction, with $\alpha < 2.5$ and a slightly earlier reionisation $z_{\rm{reion}} \approx 7$ are within $1 \sigma$ uncertainty of our results. Since the duration of reionisation also has an effect on temperatures via the ionisation front speed (see section \ref{sec:treion}, \citet{2019ApJ...874..154D}), models with harder spectra require slower reionisation histories in order to agree with both temperature and CMB measurements \citep{2019ApJ...872..101B}. As a result constraining reionisation duration will result in tighter constraints on spectral slope. We also note that spectral slopes $\alpha < 0.5$ are ruled out at $2\sigma$, as these spectra produce temperatures that are too high compared to observations, even when reionisation occurs relatively early.
\section{Conclusions}\label{sec:conc}
Using an inhomogeneous reionisation model, we have probed the full distribution of IGM temperatures, and its correlations with structure growth, the ionising background and patchy reionisation. We have begun to show how these correlations can be used to characterise the EoR, placing simultaneous constraints on the timing of reionisation and the background spectrum.
We recover thermal behaviours from other inhomogeneous reionisation models \citep{2012MNRAS.421.1969R,2008ApJ...689L..81T,2017ApJ...847...63O,2018MNRAS.477.5501K,2009ApJ...701...94F,2018arXiv181011683O}, where shortly after reionisation, there is a large scatter in the low-density end of the temperature-density relation, and a strong correlation between temperature and redshift of ionisation. By redshift z = 4, the temperature correlation shifts towards density, creating a power law temperature density relation, leaving no memory of the redshift of reionisation in the low-density IGM. Our main findings are as follows:
\begin{itemize}
\item The large initial scatter in the temperature density relation is highly correlated with the redshift of reionisation of the region, differentiating between regions near the cosmic filaments, which reionise early, and those in large voids, ionising later. Temperature measurements taken while the low-density scatter still exists contain information concerning when a region reionised, and the mean and scatter of multiple measurements can be used to constrain the reionisation history. However interpretations of these measurements will be highly dependent on the nature of the ionising sources, and the structure of the gas in the region.

\item Performing mock observations of the mean and scatter of the temperature density relation at different redshifts illustrates the potential for constraints on the EoR. Using uncertainties similar to those for recent temperature measurements \citep{2019ApJ...872..101B,2018arXiv180804367W}, we can recover the redshift of reionisation from our mock observations to $1 \sigma$ within $\Delta z_{\rm{reion}} \approx 0.5$ and background spectral slopes within $\Delta \alpha \approx 1$.

\item Comparing our suite of models to electron optical depth and temperature measurements, our modeling favours a reionisation history that finishes around $z_{\rm{reion}} = 6.8 ^{+ 0.5} _{-0.8}$ and a UV background with a power-law spectral index of $\alpha = 2.8 ^{+ 1.2} _{-1.0}$ ($1 \sigma$ uncertainties) between 912 \AA ~and 228 \AA. If we restrict our models to those with spectral slopes consistent with population II stellar sysnthesis models ($0.5 < \alpha < 2.5$), the redshift of reionisation is restricted further to $z_{\rm{reion}} = 6.9 ^{+ 0.4} _{-0.5}$.  
\end{itemize}

Knowledge of the distribution and history of temperatures within the IGM in addition to temperature at mean density will greatly improve our understanding of the EoR. while the timing of reionisation is probed by mean temperature, as discussed in section \ref{sec:mocks}, an estimate of the scatter in temperatures between different lines of sight from multiple quasar spectra, or by the large scale features of the Lyman alpha forest \citep{2018arXiv181011683O,2019arXiv190704860W}, would begin to offer information about the duration and patchiness of reionisation. This is because the distribution of temperatures shortly after reionisation is closely related to the distribution of reionisation redshifts within a volume (see Figure \ref{fig:tvd}). This is especially true for low density gas, where the distribution of temperatures is widest.

One way to further probe the temperature density relation could involve the cross-correlation of temperature measurements with the galaxy field. In this manner, information could be gained on the correlation of temperatures with density at various scales throughout reionisation. From the temperature-density plots in this work, we would expect to see the correlation on small scales flattening, or even inverting, shortly after reionisation.
\section*{Acknowledgements}
JD would like to thank James Bolton for helpful comments on the draft version of this paper.
JD is supported by the University of Melbourne under the Melbroune Research Scholarship (MRS).
This research was also supported by the Australian Research Council Centre of Excellence for All Sky Astrophysics in 3 Dimensions (ASTRO 3D), through project number CE170100013.
This work was performed on the OzSTAR national facility at Swinburne University of Technology. OzSTAR is funded by Swinburne University of Technology and the National Collaborative Research Infrastructure Strategy (NCRIS).
Parts of this work were performed on the gSTAR national facility at Swinburne University of Technology. gSTAR is funded by Swinburne and the Australian Government's Education Investment Fund.
Parameter constraints plots were generated using the \textsc{corner} python package \citep{corner}.




\bibliographystyle{mnras}
\bibliography{paper}

\begin{thebibliography}{}
\makeatletter
\relax
\def\mn@urlcharsother{\let\do\@makeother \do\$\do\&\do\#\do\^\do\_\do\%\do\~}
\def\mn@doi{\begingroup\mn@urlcharsother \@ifnextchar [ {\mn@doi@}
  {\mn@doi@[]}}
\def\mn@doi@[#1]#2{\def\@tempa{#1}\ifx\@tempa\@empty \href
  {http://dx.doi.org/#2} {doi:#2}\else \href {http://dx.doi.org/#2} {#1}\fi
  \endgroup}
\def\mn@eprint#1#2{\mn@eprint@#1:#2::\@nil}
\def\mn@eprint@arXiv#1{\href {http://arxiv.org/abs/#1} {{\tt arXiv:#1}}}
\def\mn@eprint@dblp#1{\href {http://dblp.uni-trier.de/rec/bibtex/#1.xml}
  {dblp:#1}}
\def\mn@eprint@#1:#2:#3:#4\@nil{\def\@tempa {#1}\def\@tempb {#2}\def\@tempc
  {#3}\ifx \@tempc \@empty \let \@tempc \@tempb \let \@tempb \@tempa \fi \ifx
  \@tempb \@empty \def\@tempb {arXiv}\fi \@ifundefined
  {mn@eprint@\@tempb}{\@tempb:\@tempc}{\expandafter \expandafter \csname
  mn@eprint@\@tempb\endcsname \expandafter{\@tempc}}}

\bibitem[\protect\citeauthoryear{{Anninos}, {Zhang}, {Abel}  \&
  {Norman}}{{Anninos} et~al.}{1997}]{1997NewA....2..209A}
{Anninos} P.,  {Zhang} Y.,  {Abel} T.,   {Norman} M.~L.,  1997, \mn@doi [\na]
  {10.1016/S1384-1076(97)00009-2}, \href
  {http://adsabs.harvard.edu/abs/1997NewA....2..209A} {2, 209}

\bibitem[\protect\citeauthoryear{{Becker} \& {Bolton}}{{Becker} \&
  {Bolton}}{2013}]{2013MNRAS.436.1023B}
{Becker} G.~D.,  {Bolton} J.~S.,  2013, \mn@doi [\mnras]
  {10.1093/mnras/stt1610}, \href
  {http://adsabs.harvard.edu/abs/2013MNRAS.436.1023B} {436, 1023}

\bibitem[\protect\citeauthoryear{{Becker}, {Bolton}, {Haehnelt}  \&
  {Sargent}}{{Becker} et~al.}{2011}]{2011MNRAS.410.1096B}
{Becker} G.~D.,  {Bolton} J.~S.,  {Haehnelt} M.~G.,   {Sargent} W.~L.~W.,
  2011, \mn@doi [\mnras] {10.1111/j.1365-2966.2010.17507.x}, \href
  {http://adsabs.harvard.edu/abs/2011MNRAS.410.1096B} {410, 1096}

\bibitem[\protect\citeauthoryear{{Boera}, {Becker}, {Bolton}  \&
  {Nasir}}{{Boera} et~al.}{2019}]{2019ApJ...872..101B}
{Boera} E.,  {Becker} G.~D.,  {Bolton} J.~S.,   {Nasir} F.,  2019, \mn@doi
  [\apj] {10.3847/1538-4357/aafee4}, \href
  {https://ui.adsabs.harvard.edu/abs/2019ApJ...872..101B} {872, 101}

\bibitem[\protect\citeauthoryear{{Bolton} \& {Haehnelt}}{{Bolton} \&
  {Haehnelt}}{2007}]{2007MNRAS.374..493B}
{Bolton} J.~S.,  {Haehnelt} M.~G.,  2007, \mn@doi [\mnras]
  {10.1111/j.1365-2966.2006.11176.x}, \href
  {http://adsabs.harvard.edu/abs/2007MNRAS.374..493B} {374, 493}

\bibitem[\protect\citeauthoryear{{D'Aloisio}, {McQuinn}, {Maupin}, {Davies},
  {Trac}, {Fuller}  \& {Upton Sanderbeck}}{{D'Aloisio}
  et~al.}{2019}]{2019ApJ...874..154D}
{D'Aloisio} A.,  {McQuinn} M.,  {Maupin} O.,  {Davies} F.~B.,  {Trac} H.,
  {Fuller} S.,   {Upton Sanderbeck} P.~R.,  2019, \mn@doi [\apj]
  {10.3847/1538-4357/ab0d83}, \href
  {https://ui.adsabs.harvard.edu/abs/2019ApJ...874..154D} {874, 154}

\bibitem[\protect\citeauthoryear{{Fan}, {Carilli}  \& {Keating}}{{Fan}
  et~al.}{2006}]{2006ARA&A..44..415F}
{Fan} X.,  {Carilli} C.~L.,   {Keating} B.,  2006, \mn@doi [\araa]
  {10.1146/annurev.astro.44.051905.092514}, \href
  {http://adsabs.harvard.edu/abs/2006ARA%26A..44..415F} {44, 415}

\bibitem[\protect\citeauthoryear{{Faucher-Gigu{\`e}re}, {Lidz}, {Zaldarriaga}
  \& {Hernquist}}{{Faucher-Gigu{\`e}re} et~al.}{2009}]{2009ApJ...703.1416F}
{Faucher-Gigu{\`e}re} C.-A.,  {Lidz} A.,  {Zaldarriaga} M.,   {Hernquist} L.,
  2009, \mn@doi [\apj] {10.1088/0004-637X/703/2/1416}, \href
  {http://adsabs.harvard.edu/abs/2009ApJ...703.1416F} {703, 1416}

\bibitem[\protect\citeauthoryear{Foreman-Mackey}{Foreman-Mackey}{2016}]{corner}
Foreman-Mackey D.,  2016, \mn@doi [The Journal of Open Source Software]
  {10.21105/joss.00024}, 24

\bibitem[\protect\citeauthoryear{{Furlanetto} \& {Oh}}{{Furlanetto} \&
  {Oh}}{2008}]{2008ApJ...682...14F}
{Furlanetto} S.~R.,  {Oh} S.~P.,  2008, \mn@doi [\apj] {10.1086/589613}, \href
  {http://adsabs.harvard.edu/abs/2008ApJ...682...14F} {682, 14}

\bibitem[\protect\citeauthoryear{{Furlanetto} \& {Oh}}{{Furlanetto} \&
  {Oh}}{2009}]{2009ApJ...701...94F}
{Furlanetto} S.~R.,  {Oh} S.~P.,  2009, \mn@doi [\apj]
  {10.1088/0004-637X/701/1/94}, \href
  {http://adsabs.harvard.edu/abs/2009ApJ...701...94F} {701, 94}

\bibitem[\protect\citeauthoryear{{Gaikwad}, {Srianand}, {Khaire}  \&
  {Choudhury}}{{Gaikwad} et~al.}{2018}]{2018arXiv181201016G}
{Gaikwad} P.,  {Srianand} R.,  {Khaire} V.,   {Choudhury} T.~R.,  2018,
  preprint, \href {http://adsabs.harvard.edu/abs/2018arXiv181201016G} {}
  (\mn@eprint {arXiv} {1812.01016})

\bibitem[\protect\citeauthoryear{{Greig} \& {Mesinger}}{{Greig} \&
  {Mesinger}}{2017}]{2017MNRAS.465.4838G}
{Greig} B.,  {Mesinger} A.,  2017, \mn@doi [\mnras] {10.1093/mnras/stw3026},
  \href {https://ui.adsabs.harvard.edu/abs/2017MNRAS.465.4838G} {465, 4838}

\bibitem[\protect\citeauthoryear{{Hui} \& {Gnedin}}{{Hui} \&
  {Gnedin}}{1997}]{1997MNRAS.292...27H}
{Hui} L.,  {Gnedin} N.~Y.,  1997, \mn@doi [\mnras] {10.1093/mnras/292.1.27},
  \href {http://adsabs.harvard.edu/abs/1997MNRAS.292...27H} {292, 27}

\bibitem[\protect\citeauthoryear{{Hui} \& {Haiman}}{{Hui} \&
  {Haiman}}{2003}]{2003ApJ...596....9H}
{Hui} L.,  {Haiman} Z.,  2003, \mn@doi [\apj] {10.1086/377229}, \href
  {http://adsabs.harvard.edu/abs/2003ApJ...596....9H} {596, 9}

\bibitem[\protect\citeauthoryear{{Keating}, {Puchwein}  \&
  {Haehnelt}}{{Keating} et~al.}{2018}]{2018MNRAS.477.5501K}
{Keating} L.~C.,  {Puchwein} E.,   {Haehnelt} M.~G.,  2018, \mn@doi [\mnras]
  {10.1093/mnras/sty968}, \href
  {http://adsabs.harvard.edu/abs/2018MNRAS.477.5501K} {477, 5501}

\bibitem[\protect\citeauthoryear{{Lidz} \& {Malloy}}{{Lidz} \&
  {Malloy}}{2014}]{2014ApJ...788..175L}
{Lidz} A.,  {Malloy} M.,  2014, \mn@doi [\apj] {10.1088/0004-637X/788/2/175},
  \href {http://adsabs.harvard.edu/abs/2014ApJ...788..175L} {788, 175}

\bibitem[\protect\citeauthoryear{{Luki{\'c}}, {Stark}, {Nugent}, {White},
  {Meiksin}  \& {Almgren}}{{Luki{\'c}} et~al.}{2015}]{2015MNRAS.446.3697L}
{Luki{\'c}} Z.,  {Stark} C.~W.,  {Nugent} P.,  {White} M.,  {Meiksin} A.~A.,
  {Almgren} A.,  2015, \mn@doi [\mnras] {10.1093/mnras/stu2377}, \href
  {http://adsabs.harvard.edu/abs/2015MNRAS.446.3697L} {446, 3697}

\bibitem[\protect\citeauthoryear{{McQuinn} \& {Upton Sanderbeck}}{{McQuinn} \&
  {Upton Sanderbeck}}{2016}]{2016MNRAS.456...47M}
{McQuinn} M.,  {Upton Sanderbeck} P.~R.,  2016, \mn@doi [\mnras]
  {10.1093/mnras/stv2675}, \href
  {http://adsabs.harvard.edu/abs/2016MNRAS.456...47M} {456, 47}

\bibitem[\protect\citeauthoryear{{Mesinger} \& {Furlanetto}}{{Mesinger} \&
  {Furlanetto}}{2007}]{2007ApJ...669..663M}
{Mesinger} A.,  {Furlanetto} S.,  2007, \mn@doi [\apj] {10.1086/521806}, \href
  {http://adsabs.harvard.edu/abs/2007ApJ...669..663M} {669, 663}

\bibitem[\protect\citeauthoryear{{Miralda-Escud{\'e}} \&
  {Rees}}{{Miralda-Escud{\'e}} \& {Rees}}{1994}]{1994MNRAS.266..343M}
{Miralda-Escud{\'e}} J.,  {Rees} M.~J.,  1994, \mn@doi [\mnras]
  {10.1093/mnras/266.2.343}, \href
  {http://adsabs.harvard.edu/abs/1994MNRAS.266..343M} {266, 343}

\bibitem[\protect\citeauthoryear{{Mutch}, {Geil}, {Poole}, {Angel}, {Duffy},
  {Mesinger}  \& {Wyithe}}{{Mutch} et~al.}{2016}]{2016MNRAS.462..250M}
{Mutch} S.~J.,  {Geil} P.~M.,  {Poole} G.~B.,  {Angel} P.~W.,  {Duffy} A.~R.,
  {Mesinger} A.,   {Wyithe} J.~S.~B.,  2016, \mn@doi [\mnras]
  {10.1093/mnras/stw1506}, \href
  {http://adsabs.harvard.edu/abs/2016MNRAS.462..250M} {462, 250}

\bibitem[\protect\citeauthoryear{{Nasir}, {Bolton}  \& {Becker}}{{Nasir}
  et~al.}{2016}]{2016MNRAS.463.2335N}
{Nasir} F.,  {Bolton} J.~S.,   {Becker} G.~D.,  2016, \mn@doi [\mnras]
  {10.1093/mnras/stw2147}, \href
  {http://adsabs.harvard.edu/abs/2016MNRAS.463.2335N} {463, 2335}

\bibitem[\protect\citeauthoryear{{O{\~n}orbe}, {Hennawi}  \&
  {Luki{\'c}}}{{O{\~n}orbe} et~al.}{2017a}]{2017ApJ...837..106O}
{O{\~n}orbe} J.,  {Hennawi} J.~F.,   {Luki{\'c}} Z.,  2017a, \mn@doi [\apj]
  {10.3847/1538-4357/aa6031}, \href
  {http://adsabs.harvard.edu/abs/2017ApJ...837..106O} {837, 106}

\bibitem[\protect\citeauthoryear{{O{\~n}orbe}, {Hennawi}, {Luki{\'c}}  \&
  {Walther}}{{O{\~n}orbe} et~al.}{2017b}]{2017ApJ...847...63O}
{O{\~n}orbe} J.,  {Hennawi} J.~F.,  {Luki{\'c}} Z.,   {Walther} M.,  2017b,
  \mn@doi [\apj] {10.3847/1538-4357/aa898d}, \href
  {http://adsabs.harvard.edu/abs/2017ApJ...847...63O} {847, 63}

\bibitem[\protect\citeauthoryear{{O{\~n}orbe}, {Davies}, {Luki{\'c}}, {Hennawi}
   \& {Sorini}}{{O{\~n}orbe} et~al.}{2018}]{2018arXiv181011683O}
{O{\~n}orbe} J.,  {Davies} F.~B.,  {Luki{\'c}} Z.,  {Hennawi} J.~F.,   {Sorini}
  D.,  2018, preprint, \href
  {http://adsabs.harvard.edu/abs/2018arXiv181011683O} {} (\mn@eprint {arXiv}
  {1810.11683})

\bibitem[\protect\citeauthoryear{{Planck Collaboration} et~al.,}{{Planck
  Collaboration} et~al.}{2016}]{2016A&A...594A..13P}
{Planck Collaboration} et~al., 2016, \mn@doi [\aap]
  {10.1051/0004-6361/201525830}, \href
  {http://adsabs.harvard.edu/abs/2016A%26A...594A..13P} {594, A13}

\bibitem[\protect\citeauthoryear{{Planck Collaboration} et~al.,}{{Planck
  Collaboration} et~al.}{2018}]{2018arXiv180706209P}
{Planck Collaboration} et~al., 2018, preprint, \href
  {http://adsabs.harvard.edu/abs/2018arXiv180706209P} {} (\mn@eprint {arXiv}
  {1807.06209})

\bibitem[\protect\citeauthoryear{{Poole}, {Angel}, {Mutch}, {Power}, {Duffy},
  {Geil}, {Mesinger}  \& {Wyithe}}{{Poole} et~al.}{2016}]{2016MNRAS.459.3025P}
{Poole} G.~B.,  {Angel} P.~W.,  {Mutch} S.~J.,  {Power} C.,  {Duffy} A.~R.,
  {Geil} P.~M.,  {Mesinger} A.,   {Wyithe} S.~B.,  2016, \mn@doi [\mnras]
  {10.1093/mnras/stw674}, \href
  {http://adsabs.harvard.edu/abs/2016MNRAS.459.3025P} {459, 3025}

\bibitem[\protect\citeauthoryear{{Poole}, {Mutch}, {Croton}  \&
  {Wyithe}}{{Poole} et~al.}{2017}]{2017MNRAS.472.3659P}
{Poole} G.~B.,  {Mutch} S.~J.,  {Croton} D.~J.,   {Wyithe} S.,  2017, \mn@doi
  [\mnras] {10.1093/mnras/stx2233}, \href
  {http://adsabs.harvard.edu/abs/2017MNRAS.472.3659P} {472, 3659}

\bibitem[\protect\citeauthoryear{{Puchwein}, {Haardt}, {Haehnelt}  \&
  {Madau}}{{Puchwein} et~al.}{2018}]{2018arXiv180104931P}
{Puchwein} E.,  {Haardt} F.,  {Haehnelt} M.~G.,   {Madau} P.,  2018, preprint,
  \href {http://adsabs.harvard.edu/abs/2018arXiv180104931P} {} (\mn@eprint
  {arXiv} {1801.04931})

\bibitem[\protect\citeauthoryear{{Qin} et~al.,}{{Qin}
  et~al.}{2017}]{2017MNRAS.472.2009Q}
{Qin} Y.,  et~al., 2017, \mn@doi [\mnras] {10.1093/mnras/stx1909}, \href
  {http://adsabs.harvard.edu/abs/2017MNRAS.472.2009Q} {472, 2009}

\bibitem[\protect\citeauthoryear{Raskutti}{Raskutti}{2011}]{SRthesis}
Raskutti S.,  2011, Mphil thesis, School of Physics, University of Melbourne

\bibitem[\protect\citeauthoryear{{Raskutti}, {Bolton}, {Wyithe}  \&
  {Becker}}{{Raskutti} et~al.}{2012}]{2012MNRAS.421.1969R}
{Raskutti} S.,  {Bolton} J.~S.,  {Wyithe} J.~S.~B.,   {Becker} G.~D.,  2012,
  \mn@doi [\mnras] {10.1111/j.1365-2966.2011.20401.x}, \href
  {http://adsabs.harvard.edu/abs/2012MNRAS.421.1969R} {421, 1969}

\bibitem[\protect\citeauthoryear{{Schaye}, {Theuns}, {Rauch}, {Efstathiou}  \&
  {Sargent}}{{Schaye} et~al.}{2000}]{2000MNRAS.318..817S}
{Schaye} J.,  {Theuns} T.,  {Rauch} M.,  {Efstathiou} G.,   {Sargent} W.~L.~W.,
   2000, \mn@doi [\mnras] {10.1046/j.1365-8711.2000.03815.x}, \href
  {http://adsabs.harvard.edu/abs/2000MNRAS.318..817S} {318, 817}

\bibitem[\protect\citeauthoryear{{Sobacchi} \& {Mesinger}}{{Sobacchi} \&
  {Mesinger}}{2013}]{2013MNRAS.432.3340S}
{Sobacchi} E.,  {Mesinger} A.,  2013, \mn@doi [\mnras] {10.1093/mnras/stt693},
  \href {http://adsabs.harvard.edu/abs/2013MNRAS.432.3340S} {432, 3340}

\bibitem[\protect\citeauthoryear{{Songaila} \& {Cowie}}{{Songaila} \&
  {Cowie}}{2010}]{2010ApJ...721.1448S}
{Songaila} A.,  {Cowie} L.~L.,  2010, \mn@doi [\apj]
  {10.1088/0004-637X/721/2/1448}, \href
  {http://adsabs.harvard.edu/abs/2010ApJ...721.1448S} {721, 1448}

\bibitem[\protect\citeauthoryear{{Theuns}, {Schaye}, {Zaroubi}, {Kim},
  {Tzanavaris}  \& {Carswell}}{{Theuns} et~al.}{2002}]{2002ApJ...567L.103T}
{Theuns} T.,  {Schaye} J.,  {Zaroubi} S.,  {Kim} T.-S.,  {Tzanavaris} P.,
  {Carswell} B.,  2002, \mn@doi [\apjl] {10.1086/339998}, \href
  {http://adsabs.harvard.edu/abs/2002ApJ...567L.103T} {567, L103}

\bibitem[\protect\citeauthoryear{{Trac}, {Cen}  \& {Loeb}}{{Trac}
  et~al.}{2008}]{2008ApJ...689L..81T}
{Trac} H.,  {Cen} R.,   {Loeb} A.,  2008, \mn@doi [\apjl] {10.1086/595678},
  \href {http://adsabs.harvard.edu/abs/2008ApJ...689L..81T} {689, L81}

\bibitem[\protect\citeauthoryear{{Upton Sanderbeck}, {D'Aloisio}  \&
  {McQuinn}}{{Upton Sanderbeck} et~al.}{2016}]{2016MNRAS.460.1885U}
{Upton Sanderbeck} P.~R.,  {D'Aloisio} A.,   {McQuinn} M.~J.,  2016, \mn@doi
  [\mnras] {10.1093/mnras/stw1117}, \href
  {http://adsabs.harvard.edu/abs/2016MNRAS.460.1885U} {460, 1885}

\bibitem[\protect\citeauthoryear{{Verner}, {Ferland}, {Korista}  \&
  {Yakovlev}}{{Verner} et~al.}{1996}]{1996ApJ...465..487V}
{Verner} D.~A.,  {Ferland} G.~J.,  {Korista} K.~T.,   {Yakovlev} D.~G.,  1996,
  \mn@doi [\apj] {10.1086/177435}, \href
  {http://adsabs.harvard.edu/abs/1996ApJ...465..487V} {465, 487}

\bibitem[\protect\citeauthoryear{{Walther}, {O{\~n}orbe}, {Hennawi}  \&
  {Luki{\'c}}}{{Walther} et~al.}{2018}]{2018arXiv180804367W}
{Walther} M.,  {O{\~n}orbe} J.,  {Hennawi} J.~F.,   {Luki{\'c}} Z.,  2018,
  preprint, \href {http://adsabs.harvard.edu/abs/2018arXiv180804367W} {}
  (\mn@eprint {arXiv} {1808.04367})

\bibitem[\protect\citeauthoryear{{Wu}, {McQuinn}, {Kannan}, {D'Aloisio},
  {Bird}, {Marinacci}, {Dav{\'e}}  \& {Hernquist}}{{Wu}
  et~al.}{2019}]{2019arXiv190704860W}
{Wu} X.,  {McQuinn} M.,  {Kannan} R.,  {D'Aloisio} A.,  {Bird} S.,  {Marinacci}
  F.,  {Dav{\'e}} R.,   {Hernquist} L.,  2019, arXiv e-prints, \href
  {https://ui.adsabs.harvard.edu/abs/2019arXiv190704860W} {}

\bibitem[\protect\citeauthoryear{{Wyithe} \& {Loeb}}{{Wyithe} \&
  {Loeb}}{2003}]{2003ApJ...586..693W}
{Wyithe} J.~S.~B.,  {Loeb} A.,  2003, \mn@doi [\apj] {10.1086/367721}, \href
  {http://adsabs.harvard.edu/abs/2003ApJ...586..693W} {586, 693}

\makeatother
\end{thebibliography}



\appendix

\section{Numerical convergence}\label{sec:gridres}
This section explores the effects of numerical factors in our model. The grid resolution of \textsc{Meraxes} will be investigated, as well as the parameters used in our DE solver, namely the convergence threshold that we use to determine the timestep throughout the model.

\begin{figure}
\includegraphics[width=\linewidth]{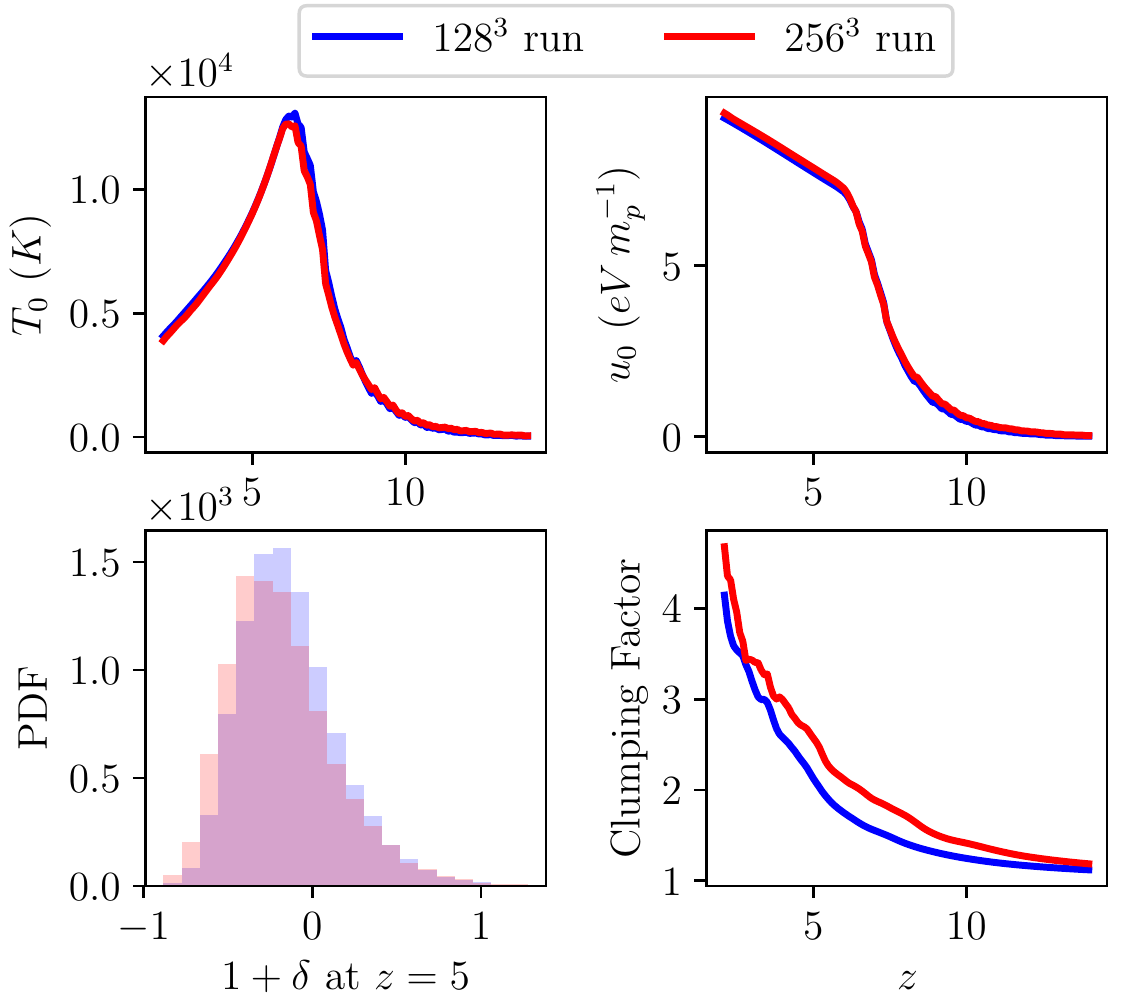}
\caption{Comparison of runs on $128^3$ and $256^3$ grids. Top left: Temperature at mean density. Top right: photo-ionisation energy at mean density. Bottom left: density distributions. Bottom right: clumping factors above the grid scale $\frac{\langle \delta^2 \rangle}{\langle \delta \rangle ^2}$, distributions of density and clumping will change, but the thermal state of gas at a particular density will be unchanged}
\label{fig:resolution}
\end{figure}

For computational reasons, all of model runs were performed using $128^3$ grids in \textsc{Meraxes}. However to ensure inhomogeneities below this scale do not greatly affect our results, we examine results from one run on a $256^3$ grid. As shown in Figure \ref{fig:resolution}, while the overall clumping factor has increased, the temperatures and photo-heating energy at mean density are largely unaffected. This is due to the fact that we follow the gas at the mean density of each voxel, rather than the voxel as a whole. This will produce a very similar temperature-density relation regardless of grid resolution; The distribution of densities will change, but the temperature at any particular density will remain the same.

\begin{figure}
\includegraphics[width=\linewidth]{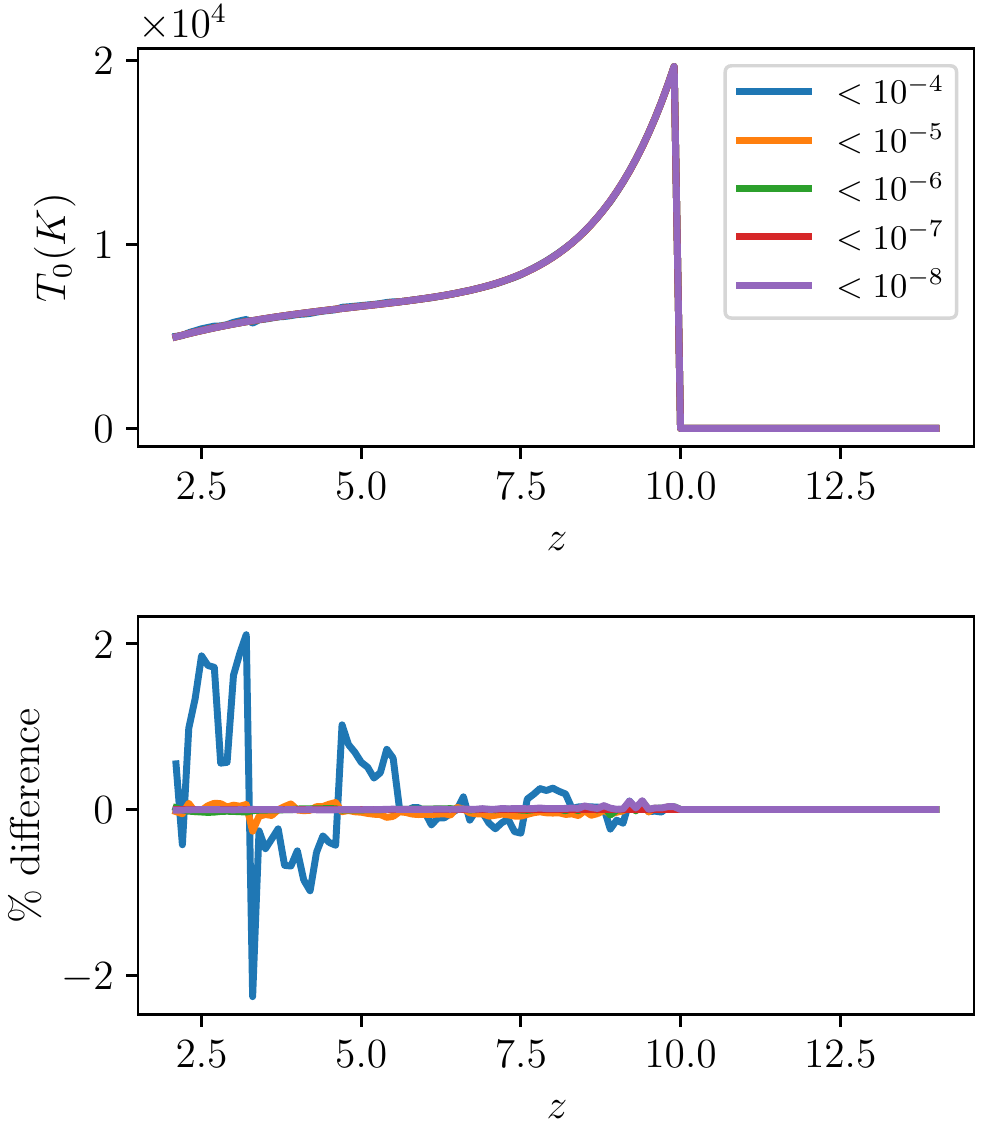}
\caption{Temperatures for single voxel runs with changes in DE solver threshold, where the fiducial threshold for accepting the next timestep is is $|\tilde{X}_{e,k+1} - \tilde{X}_{e,k}| < 10^{-8}$ to $10^{-4}$. Tightening this threshold will only affect results by a fraction of a percent.}
\label{fig:deconv}
\end{figure}

We have altered the convergence conditions in our differential equation solver to test for convergence. We change the conversion threshold between $|\tilde{X}_{e,k+1} - \tilde{X}_{e,k}| < 10^{-8}$ and $10^{-4}$. The results of these changes for an individual voxel that ionised at $z=10$ are presented in Figure \ref{fig:deconv}. The weakest threshold gives a maximum difference in temperatures of $\approx 2 \%$, and our fiducial threshold of $10^{-6}$ is negligibly different from the strictest thresholds. We are therefore confident that our choice of differential equation solver parameters does not affect our results.

\section{Variation with Ionising Flux Amplitude}\label{sec:gammachange}
\begin{figure}
  \centering
  \includegraphics[width=\linewidth]{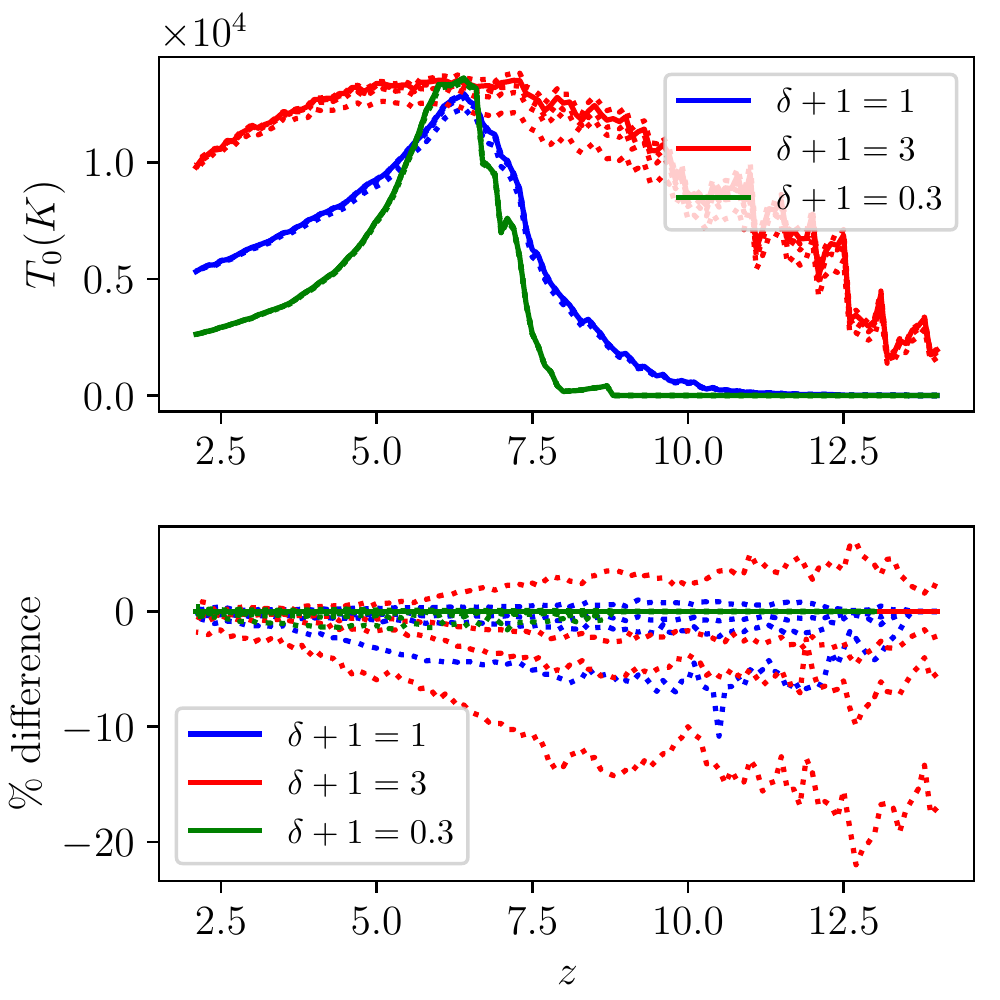}
  \caption{Temperature at mean density versus redshift for each ionising flux amplitude tested. Blue lines show results for voxels at mean density, red lines show overdense voxels $\delta + 1 = 3$, green lines show underdense voxels $\delta + 1 = 0.3$. Dotted lines from top to bottom show models where the ionising flux amplitude has been altered by a factor of 10,2,0.5 and 0.1 respectively. Top panel: Temperature at mean density. Bottom panel: percentage difference from the base model.}
\label{fig:J21means}
\end{figure}

Previous simulations have found little dependence of the long term thermal state on the amplitude of the ionising background, as long as it is strong enough to maintain an ionised IGM \citep{1997MNRAS.292...27H,2009ApJ...701...94F}. This result is verified in our model, as we can see no change in the long term thermal state when we artificially increase or decrease the amplitude of the ionising background post-reionisation, keeping our other parameters constant. The ionisation rate in a voxel is directly proportional to the amplitude, however the number of ionisations that actually occur is limited by the recombination rate, which is negligibly changed for all highly ionised states. As a result the long term cooling is largely unaffected at most densities. Figure \ref{fig:J21means} shows the mean temperature at mean density, a third of mean density, and three times mean density, for the three amplitudes tested.

It is important to note that the ionising flux amplitude has little effect on temperature only when there is enough ionising flux to maintain the ionised state in a region. If the amount of ionising flux is low enough such that it is comparable to or lower than the recombination rate, there will be a significant drop in temperature as the gas recombines. Furthermore, the long term temperatures will decrease, due to the ionisation rate falling below the recombination rate for a fully ionised state; this causes the ionisation equilibrium state to shift to a more neutral state, lowering the rate of ionisations and recombinations, and therefore the photo-heating rate near equilibrium, as the cooling rate is now limited by ionisations, rather than recombinations.

In our simulation, there is enough ionising flux with the fiducial values from \textit{\textsc{Meraxes}} to maintain a highly ionised state in the vast majority of the simulated volume, so a significant drop in temperature is observed only for highly dense regions. Voxels at mean density will only show a 5\% decrease in temperatures when the ionising flux is decreased by a factor of 10. Differences in temperature will also diminish over time, as the gas cools to its thermal asymptote.

Regions near early galaxies that have their star formation suppressed or move between voxels can also show this behaviour, however in this case the temperature drop is also usually very short-lived, as nearby HII bubbles expand to re-heat the voxel, washing out any memory of previous ionisation events.


\bsp	
\label{lastpage}
\end{document}